# Origin of the Chemical Elements

T. Rauscher[1] , A. Patkós[2]

[13]*Department of Physics, University of Basel, Basel, Switzerland*

[2]*Department of Atomic Physics, Eötvös Loránd University, Budapest, Hungary*

Summary:     This chapter provides the necessary background from astrophysics, nuclear, and particle physics to understand the cosmic origin of the chemical elements. It reflects the year 2009 state of the art in this extremely quickly developing interdisciplinary research direction. The discussion summarizes the nucleosynthetic processes in the course of the evolution of the Universe and the galaxies contained within, including primordial nucleosynthesis, stellar evolution, and explosive nucleosynthesis in single and binary systems.

# CONTENTS

# 1    INTRODUCTION

Chemical elements are central for the existence of life and the richness and variety of our environment. Therefore, one of the basic questions concerns the origin of the chemical elements. The answer is complex because it relies on dynamical processes from elementary particles and nuclei to stars and galaxies. An interdisciplinary effort of various fields of science achieved considerable progress in this direction of research. The present review summarizes the state of knowledge obtained mainly from particle and nuclear physics, astrophysics and astronomy.

In Sections 2 and 3 we concentrate on the two most important information sources concerning the earliest history of the Universe, i.e. the cosmic microwave background radiation and the primordial synthesis of the nuclei of the lightest chemical elements. Our aim is to describe, in the simplest qualitative terms, the empirical facts and the way their interpretation is connected with the physics of the epoch immediately following the Big Bang. It should become clear that the structures observed today on the largest distance scales reflect the nature of the quantum fluctuations of the earliest period. Moreover, nuclear physics combined with the basic facts of cosmology provide a perfect account of the primordial abundance of the lightest nuclei. In Section 4 the production mechanism of the elements will be discussed as they occur in the different stages of stellar evolution. Explosive events occurring in binary stellar systems and their roles in the nucleosynthesis are discussed in Section 5. The concluding Section 6 is devoted to the description and the interpretation of the abundance of chemical elements in the Sun and in the Galaxy. This includes abundance determinations from astronomical observations as well as from the analysis of presolar grains. The experimental methods to determine abundances and to study the nuclear physics relevant for nucleosynthesis processes are outlined. Finally, the basic ideas of Galactic Chemical Evolution are laid out, which combines all the knowledge concerning production and distribution of nuclides to a grander picture. The chapter is completed by a list of references, where textbooks and review articles appear alongside the relevant original publications.

# 2    CREATION AND EARLY EVOLUTION OF MATTER IN THE UNIVERSE

## 2.1    Evolution of the energy density in the early Universe

The basic question addressed when investigating the history of the Universe as a whole in the framework of modern physics is the following: Why do we see something instead of detecting nothing? It originates from the common wisdom that any isolated system after long enough evolution will reach thermal equilibrium, characterized by a homogenous structureless distribution of its energy. Nearly 14 billion years after the Big Bang one observes the presence of complicated hierarchical structures on all scales, starting from the subnuclear world, through chemical elements, and up to the scale of galaxy clusters. This section will review our present understanding of how the structured evolution of the Universe could be sustained for a time more than 60 orders of magnitude longer than the characteristic time scale of the particle physics processes present at the moment of its 'birth'.

The information concerning the constitution of the early Universe has increased tremendously during the past decade, mainly due to improved observations of the **Cosmic Microwave Background Radiation (CMBR)**. The most important cosmological parameters (the total energy density, the part contained in baryonic matter, the part of non-baryonic dark matter, other components, etc.) have been determined with percent level accuracy as a result of projects completed in the first decade of the 21st century and now appear in tables of fundamental physical data (Amsler 2008).

### 2.1.1    *Observations of CMBR*

The existence of CMBR was predicted by Alpher *et al.* (1948) as a direct consequence of the Hot Big Bang Universe of Gamow (Lamarre and Puget 2001). It was discovered by Penzias and Wilson (1965). It originates from the combination of the once free electrons and protons into neutral atoms when the temperature of the Universe dropped below $kT = 13.6$ eV (the ionization energy of the H-atom, i.e., $T = 1.58 \times 10^5$ K) to nearly 1 eV ($1.16 \times 10^4$ K). (Note: In certain branches of physics it is customary to express temperature in eV units through the equation $E = kT$. The conversion is given by 1 eV corresponds to $1.16045 \times 10^4$ K.) After the recombination, the Universe became transparent to this radiation,  which at present reaches the detectors with a redshift determined by the kinematics of the expansion of the Universe (Lamarre and Puget 2001). It appears as a perfect thermal radiation with Planckian power distribution over more than three decades of frequency, having a temperature of $T = 2.725 \pm 0.001$ K.

The first quantitative evidence for the temperature anisotropy of CMBR was provided by the **COBE (Cosmic Background Explorer)** satellite in 1992. The angular resolution of its detectors was 7°. This enabled the collaboration to determine the first 20 multipole moments of the fluctuating part of CMBR beyond its isotropic component. It has been established that the degree of anisotropy of CMBR is one part in one hundred thousand ($10^{-5}$). There are two questions of extreme importance related to this anisotropy:

1. Is this anisotropy the origin of the hierarchical structure one observes today in the Universe?
2. What is the (micro)physical process behind this anisotropy?

We shall return to the answer to the first question in Subsection 2.4. To the second question, we will briefly outline the answer below.

Following the success of the COBE mission several more refined (ground based and balloon) measurements of the CMBR fluctuations were realized between 1998 and 2001. An angular resolution of about one degree has been achieved, which was further refined to the arc-minute level by the satellite mission **Wilkinson Microwave Anisotropy Probe (WMAP).** The combined efforts of these investigations allowed the determination of the multipole projection of CMBR on the sky up to angular moments $l = 2000$. The fluctuation information extracted until 2007 is presented in FIGURE 1 with $l_{max} = 2000$. One easily recognizes the presence of three pronounced maxima in this figure (possible additional, weaker maxima are discussed further below).

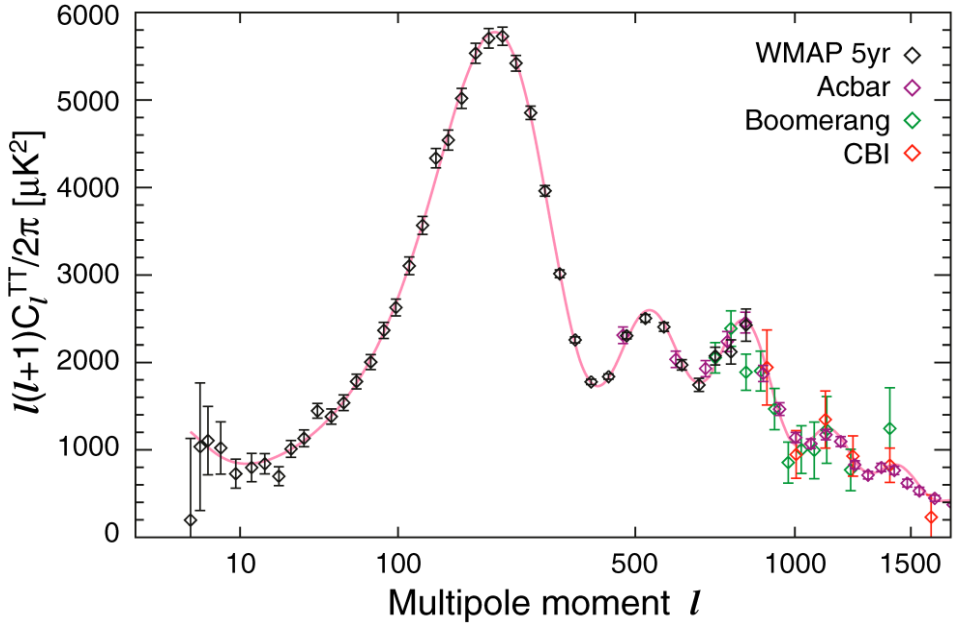

FIGURE 1. Multipole fluctuation strength of the cosmic microwave background radiation as a function of the spherical harmonic index $l$. The location and the height of the first minimum favors a spatially flat Universe, while the level of the fluctuations in the higher multipoles ($l > 400$) indicates the presence of a low-density baryonic component ($< 5\%$). The measurements cover already the damping region ($l > 1000$). WMAP data are displayed together with results of earlier balloon observations. [Reprinted from Nolta *et al.* 2009 with kind permission of the first author, the WMAP Science Team, and AAS.]

Another important characteristic of the CMBR anisotropy is its spectral power distribution. The measured distribution is nearly scale invariant; it is the so-called Zel'dovich-Harrison spectrum (see Peebles 1993). This means that every unit in the logarithm of the wave number contributes almost equally to the total power.

The small-amplitude and almost scale-invariant nature of the fluctuation spectra, described above, reflects the very early fluctuations of the gravitational field. First of all, one has to emphasize that the coupled electron-proton-photon plasma near recombination was oscillating in a varying gravitational field (Hu 2001). Where the energy density was higher the plasma experienced the effect of a potential well, and the radiation emerging from this region was hotter than average. On the contrary, diminutions of the energy density led to a colder emission. Still, an observer located far from the sources detects lower temperature from denser sources due to the Sachs-Wolfe effect (Peebles 1993). In any case, the CMBR anisotropy actually traces the inhomogeneity of the gravitational potential (or total energy density) in the era of recombination.

Thomson scattering of the anisotropic CMBR on the ionized hot matter of galaxy clusters and galaxies results in roughly 5% linear polarization of CMBR. Its presence in CMBR was first detected by the Degree Angular Scale Interferometer (DASI) experiment (J. Kovac 2002). Starting from 2003 the WMAP experiment measured also the temperature-polarization cross correlation jointly with the temperature-temperature correlation. The significance of this type of measurement is obvious since the presence of ionized gases corresponds to the beginning of the epoch of star formation.

### 2.1.2    Inflationary interpretation of the CMBR

A unique particle physics framework has been proposed which can account for the energy density fluctuations with the characteristics found in CMBR. One conjectures that the large-scale homogeneity of the Universe is due to a very early period of exponential inflation in its scale (Peebles 1993).

One assumes that during the first era after the Big Bang the size of the causally connected regions (the horizon) remained constant, while the global scale of the Universe increased exponentially. This is called inflationary epoch. The wavelength of any physical object is redshifted in proportion with the global scale. Therefore, at a certain moment fluctuations with a wavelength bigger than the horizon were 'felt' as constant fields and did not influence anymore the gravitational evolution of the matter and radiation at smaller length scale.

The inflationary period in the evolution of the Universe ended at about $10^{-32}$ s after the Big Bang. At this moment the constant ordered potential energy density driving the inflation decayed into the particles observed today. Some of them may have belonged to a more exotic class, which can contribute to the violation of the matter-antimatter symmetry if they exhibit sufficiently long lifetimes (see Section 2.2). The rate of expansion of the horizon in the subsequent radiation- and matter-dominated eras was always faster than the global expansion of the Universe (see Figure 2). Radiation-dominated means that the main contribution to the energy density comes from massless and nearly massless particles with much lower rest mass energy than the actual average kinetic energy. Therefore, the long-wavelength fluctuations having left during inflation continuously re-entered the horizon and their gravitational action was 'felt' again by the plasma oscillations. The first maximum of the CMBR multipole moments corresponds to the largest wavelength fluctuations that were just entering the horizon in the moment of the emission of CMBR.

Since during its evolution beyond the horizon, any dynamical change in the fluctuation spectra was causally forbidden, the fluctuating gravitational field experienced by the recombining hydrogen atoms was directly related to the fluctuation spectra of the inflationary epoch, determined by the quantum fluctuations of the field(s) of that era. This observation leads promptly to the conclusion that the spectra should be very close to the Zel'dovich-Harrison type. Detailed features of the power spectra seem to effectively rule out some of the concurrent inflationary models.

Also the simplest version of the field-theoretical realization of inflation predicts a total energy density very close to the critical density $\rho_c$, which separates the parameter region of a recollapsing Universe from the region where a non-accelerating expansion continues forever. Such a Universe is spatially flat. In the apparently relevant case of accelerated expansion the borderline is shifted and universes somewhat above the critical densities might expand with no return. It is customary to measure the density of a specific constituent of the universe in proportion to the critical density: $\Omega_i = \rho_i/\rho_c$.

An important prediction of the inflationary scenario for the origin of CMBR anisotropy is a sequence of maxima in the multipole spectrum (Hu 2001). The latest results (see FIGURE 1) confirm the existence of at least two further maxima, in addition to the main maximum known before. The new satellite-based CMBR observations by the European satellite **PLANCK** launched in May 2009 will improve the accuracy of the deduced cosmological parameters to 0.5% and determine the multipole projection of the anisotropy up to angular momentum l ~ 2500.

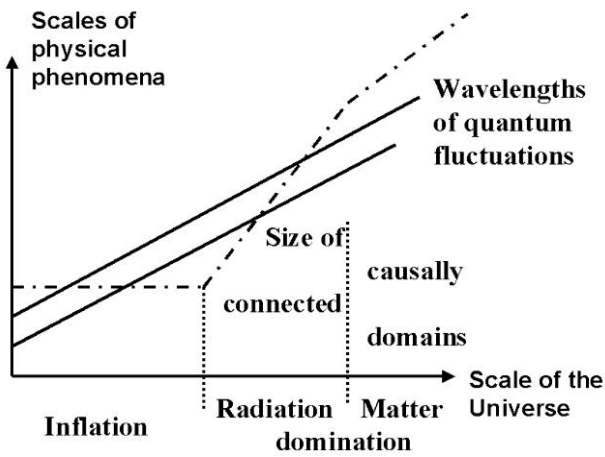

FIGURE 2. Variation of the characteristic length scales during the history of the Universe. On both axes the logarithm of the corresponding length is measured. The wavelengths of physical phenomena (full lines) grow linearly with the scale parameter of the Universe. The size of the causally connected domains (dot-dashed line) stagnates during the exponential growth (inflation), whereas it increases faster than the length scale of the Universe later, i.e. quadratically in the radiation era and with 3/2 power under matter domination.

The positions and the relative heights of these maxima allow the determination of the relative density of baryonic constituents among the energy carriers. The increased level of accuracy leads to the conclusion that the baryonic matter (building up also the nuclei of all chemical elements) constitutes no more than 5% of the energy content of the Universe. This conclusion agrees very convincingly with the results of the investigation of the primordial abundance of the light chemical elements to be described in detail in Section 3. These facts lead us to the unavoidable conclusion that about 95% of the energy content of the Universe is carried by some sort of **non-baryonic matter**. (More accurate numbers will be given at the end of this section). The discovery of its constituents and the exploration of its extremely weak interaction with ordinary matter is one of the greatest challenges for the scientific research in the 21st century.

### *2.1.3  Dark Matter: indications, candidates and signals*

Beyond CMBR, growing evidence is gathered on a very wide scale for the existence of an unknown massive constituent of galaxies and galaxy clusters. It is tempting to follow a unified approach describing the "missing gravitating mass" from the galactic to cosmological scale (i.e. from a few tens of kpc to Mpc, with $1\,pc = 3.26$ light-years $= 3.0856 \times 10^{13}$ km). In this subsection we shortly review the main evidence already found and the ongoing experimental particle physics efforts for direct detection of the Dark Matter constituents.

First hints for some sort of gravitating Dark Matter below the cosmological scale came from galactic rotation curves (some tens of kpc), then from gravitational lensing (up to 200 kpc), and from the existence of hot gas in galaxy clusters. The anomalous flattening of the rotation curves of galaxies has been discovered in the 1970s. Following Kepler's law one expects a decrease of the orbiting velocity of all objects (stars as well as gas particles) with increasing distance from the galactic center. Instead, without exception a tendency for saturation in the velocity of bright objects in all studied galaxies is observed. The simplest explanation is the existence of an enormous dark matter halo. Since the velocity measurements are based on the 21cm hydrogen hyperfine radiation, they cannot trace the galactic gravitational potential farther than a few tens of kiloparsecs. Therefore with this technique only the rise of the galactic dark matter (DM) haloes can be detected but one cannot find their extension.

Dark supermassive objects of galactic cluster size are observable by the lensing effect exerted on the light of farther objects located along their line of sight. According to General Relativity the light of distant bright objects (galaxies, quasars, bursts of gamma rays, for short: GRBs) is bent by massive matter located between the event and the observer along the line of sight. Multiple and/or distorted images arise which allow an estimate of the lensing mass. The magnitude of this effect, as measured in the Milky Way, requires even more DM out to larger distances than it was called for by the rotation curves (Adelmann-McCarthy *et al.* 2005).

The large scale geometry of the galactic DM profile semi-quantitatively agrees with results of Newtonian many-body simulations, though there are definitely discrepancies between the simulated and observed gravitating densities at shorter distances. Interesting propositions were put forward by Milgrom to cure the shorter scale deviations with a Modified Newtonian Dynamics (MOND) (reviewed by M. Milgrom 2008).

Gravitational lensing is combined with X-ray astronomy and can trace the separation of bright and dark matter, occurring when two smaller galaxies collide. The motion of the radiating matter is slowed down more than that of the DM components. As a consequence, the centers of the lensing and X-ray images are shifted relative to each other. A recent picture taken by the Chandra X-ray Telescope is considered as the first direct evidence for the existence of DM on the scale of galaxy clusters (D. Clowe *et al.* 2006).

Another way to estimate the strength of the gravitational potential in the bulk of large galaxy clusters is offered by measuring spectroscopically the average kinetic energy (e.g. the temperature) of the gas. One can relate the very high temperature values (about. $10^8$ K) to the depth of the gravitational potential assuming the validity of the virial theorem for the motion of the intergalactic gas particles. Without the DM contribution to the binding potential the hot gas would have evaporated long time ago.

There are three most popular DM candidates which could contribute to the explanation of the above wealth of observations. Historically, faint stars/planetary objects constituted of baryonic matter were invoked first, with masses smaller than 0.1 solar mass (this is the mass limit minimally needed for nuclear burning and the subsequent electromagnetic radiation). The search for Massive Compact Halo Objects (MACHOs) was initiated in the early 1990s based on the so-called microlensing effect – a temporary variation of the brightness of a star when a MACHO crosses the line of sight between star and observer. This effect is sensitive to all kind of dark matter, baryonic or non-baryonic. The very conservative combined conclusion from these observations and some theoretical considerations is that at most 20% of the Galactic Halo can be made of stellar remnants (Alcock *et al.* 2000).

Complementary to this astronomy-based proposition elementary particle physics suggests two distinct non-baryonic "species" which originate from the extreme hot period of the universe and therefore could be present nearly homogeneously on all scales. *Axion*s are hypothetical particles of small ($10^{-(3-6)}$ eV/c$^2$) rest mass energy. They were introduced (Peccei and Quinn 1977) for the theoretical explanation of the strict validity of the symmetry of strong interactions (QCD) under the combined application of space- and charge reflections (CP-invariance). Although they are very light their kinetic energy is negligible, since they are produced in non-thermal processes. This way they represent the class of Cold Dark Matter. The parameter space was and is thoroughly searched for axions in all particle physics experiments of the last two decades. The presently allowed mass range is close to the limit of the astrophysical significance of these particles.

The most natural DM candidates from particle physics are Weakly Interacting Massive Particles (WIMPs). Assuming that the thermal abundance of the WIMPs is determined by the annihilation and pair-production processes with themselves, one can estimate their present density as a function of the annihilation cross-section. It is quite remarkable (and is even qualified sometimes as "WIMP miracle") that, using cross-sections typical for the supersymmetric extension of the standard particle physics model, just the gravitating density missing on the cosmological scale is found. By this coincidence one identifies WIMPs with the lightest stable supersymmetric particle (called neutralino). Its mass is on the scale of heavy nuclei, usually one assumes it to equal the mass of the tungsten atom. It would constitute pressureless cold dark matter, with a density calculable by analyzing its decoupling from thermal equilibrium.

An important milestone in the WIMP-story will be reached once the Large Hadron Collider (LHC) at CERN begins working. The available energy covers the expected mass range of the most popular variants of supersymmetric extensions. Currently, extensive strategies are worked out for the identification of prospective new massive particles to be observed at LHC, along with their cosmologically motivated counterparts (Baltz *et al.* 2006).

A positive identification at CERN would give new impetus to the underground direct searches WIMPs. These look for heat deposition by particles arriving from the nearest galactic neighbourhood into cryogenic detectors, well isolated from any other type of heat exchange. At present, only a single such experiment, i.e. the Dark Matter (DAMA) experiment in Gran Sasso, Italy, has reported a positive signal in the DM particle search. Already for more than five years, a seasonal variation in the heat deposition rate is observed, which may be caused by the DM particle flux variation along the orbit of the Earth (Bernabei, 2003).

Although the analysis of CMBR excludes the domination of hot dark matter, i.e. relativistic weakly interacting particles, like light neutrinos, there still exists a plethora of more exotic propositions for the constituents of dark matter, not yet accessible for experimental verification, like primordial black holes, non-thermal WIMPzillas, and the so-called Kaluza-Klein excitations of higher dimensional theories.

### 2.1.4 Dark energy, the accelerating Universe, and the problem of distance measurements

The expansion of the Universe is conventionally characterized by the Hubble law, stating that cosmological objects uniformly recede from the observer with a velocity proportional to their distance. The proportionality factor $H$ has not remained constant during the evolution of the Universe, the rate of change being characterized by the deceleration parameter $q_0 = dH^{-1}/dt - 1$ (Peebles 1993).

The deceleration can be probed by distance measurements using type Ia supernovae. As explained in Section 5.3 these very energetic cosmic events occur in a rather narrow mass range of compact objects, with a minimal scatter in their energy output or *luminosity L*, which determines also the energy flux $F$ reaching the observer at distance $d_L$ called luminosity distance. Therefore type Ia supernovae are standardizable light sources (standard candles), their light curves can be transformed into a universal form.

Standard candles are important tools to measure astronomical distances. Knowing the luminosity (i.e. energy output) of an object, it is straightforward to calculate its distance by the observed brightness which drops with the inverse square of the distance $1/r^2$. The advantage of using SN Ia is that they are outshining all the stars in a regular galaxy and thus can be seen and studied over vast distances.

More recently, the method of measuring distances with SN Ia has acquired some fame by showing that the Universe is expanding faster at large distances than expected by the standard cosmological model (Perlmutter *et al.* 1999, A.G. Riess *et al.* 1998, Leibundgut 2001a,b).

The luminosities of 42 SN Ia were analysed in these pioneering publications as a function of their redshift. The survey comprised objects with redshift $z \le 1$ which corresponds to an age $\le 10$ Gy (gigayears). Assuming that the absolute magnitude of these objects is independent of the distance (excluding evolution effects) the apparent luminosities were detected on the average 60% fainter than expected in a Universe, whose energy density is dominated by non-relativistic matter. A number of data points with $z \le 0.7$ is displayed in Figure 3, all having positive deviation for the difference of the apparent and absolute luminosities, $m-M$ (note that the fainter is a source the larger is its magnitude). The simplest interpretation is to assume the scattering of the light on its way from the source by some sort of 'dust' (full line) leading to objects which are fainter than foreseen. Less conventional is to assume positive acceleration of the global expansion. (An accelerating source is located farther away, and will appear to be fainter at a certain $z$ than expected in standard cosmology.)

The quantitative argument is based on relating the measurements to the deceleration parameter defined above. In fact the luminosity distance of an object at red-shift $z$ can be expressed as an integral of an expression of the varying Hubble-parameter $H(z)$ on the interval *(0,z)*, where *z=0* corresponds to the observer's position today. When the red-shift is not too large, one can expand this integral into a series of $z$ and arrive in the first approximation at a simple linear relation expressing Hubble's law of the dependence of the luminosity distance on the red-shift. Its first non-linear correction involves the deceleration parameter: $d_L = (c/H_0)z[1+(1-q_0)z/2+...]$. Dust absorption diminishes the source brightness irrespective the value of $z$. On the other hand, the presence of both matter and a cosmological constant will change the sign of the deceleration parameter with $z$.

In the past decade several projects contributed to the luminosity distance measurements and by now the list includes over 200 events. Specifically with the help of the Hubble telescope 13 new SnIa were found with spectroscopically confirmed redshifts exceeding $z=1$ and at present the full sample contains already 23 $z>1$ objects (Riess *et al.* 2007). Such objects most strongly influence the value of the deceleration parameter. A combined analysis of all SnIa data yields a deceleration parameter value of -0.7±0.1 (Kowalski *et al.* 2008). Its negative value signals an accelerating expansion rate at distance scales comparable to the size of the Universe.

A nonzero cosmological constant $\Lambda$ in the equations describing the dynamics of the Universe can account for such an acceleration. The cosmological constant is related to a vacuum energy density ($\rho$) characterized by negative pressure (equation of state *p=wρ,* with *w=-1*). Nowadays, the more general term '**dark energy**' is used for the hypothetical agent of such an accelerating effect (acceleration requires *w<-1/3*).

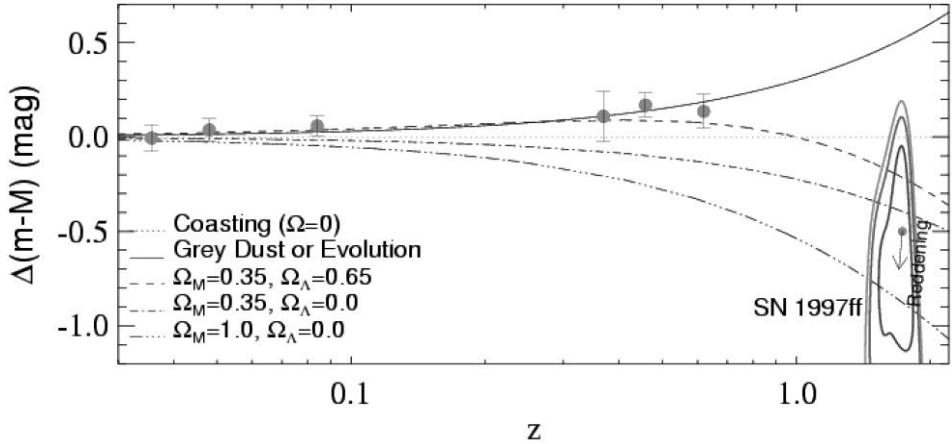

FIGURE 3. Variation of the difference of the observed ($m$) and absolute ($M$) luminosity for the SN Ia with redshift $z$, measured in a special astronomical unit, called *magnitude* (from Riess *et al.* 2001). The zero level corresponds to supernovae in an empty ($\Omega = 0$) Universe. A positive difference signals sources which are fainter than expected. A brighter (negative) value is naturally interpreted as the decelerating action of the gravitational force exerted on the source by a nonzero mass energy ($\Omega_M \neq 0$). See text for models corresponding to the different curves possibly producing positive values. [Reprinted from Riess *et al.* 2001 with kind permission of the first author and AAS.]

The fraction of the total energy density stored in $\Lambda$ is not constant in time. Although $\Lambda$ is constant by definition, the energy densities in the radiation and matter components are varying respectively as quartic and cubic inverse powers of the distance scale of the Universe. The early Universe until the decoupling of the photons is radiation-dominated and the $\Lambda$ energy density is negligible. Initially, the matter density (including non-baryonic dark matter) is dominating after decoupling, exerting a conventional decelerating effect on the motion of cosmological objects. Because of the reduction of the matter density with the expansion of the Universe, its contribution to the total energy density (and total $\Omega = \Omega_M + \Omega_\Lambda$) may become smaller than the one of $\Lambda$ at a given point in time. How early this crossover happens depends on the absolute value of $\Lambda$ which is not constrained by any current theory. From this point of view it is remarkable that $\Omega_M$ and $\Omega_\Lambda$ are of the same order of magnitude in the Universe at present. Because of this and the rather small value of $\Lambda$ its impact on the cosmic expansion can only be detected over large distances, i.e. when studying distances of objects with large redshift $z$. In the distant future, the repulsive action of a non-zero $\Lambda$ will more and more dominate. Without any additional effect, this leads to a ``Big Rip'' scenario in which smaller and smaller volumes become causally isolated because the repulsion will be pushing everything apart faster than the speed of light.

The different fits in Figure 3 correspond to different matter-dark energy compositions. Without cosmological constant, the deviation is always negative but its rate of decrease depends on $\Omega_M$. The two lower curves show cases ($\Omega_M = 0.35$ and 1.00, respectively) which do not fit the measurements at all. Evidently, if the measurements are correct, we need either some sort of dust or dark energy.

Because of the important consequences of these observations on cosmology, astronomers seriously investigate possible alternatives or data biases, such as possible effects of the evolution of SN Ia objects (deviation from the 'standard candle' behavior at low metallicity), effects of light absorption by the galaxy clusters hosting the supernovae and by the intergalactic dust, lensing effects, etc. Although the actual dust content in the line of sight is not well determined, dust is not a problem in recent observations because astronomers make use of an empirical relation between the width of the SN Ia lightcurve (between rise and decay) and its absolute magnitude. By studying samples of closer SN Ia it was found that more energetic, brighter explosions also lead to a wider lightcurve. This is called the **Phillips relation**. It is not affected by dust absorption and the only assumption entering is that it is universally valid for all SN Ia. Thus, knowing the easily determinable lightcurve width, the absolute magnitude can be derived and the distance calculated by comparison to the observed relative magnitude of the explosion.

The best fit rather indicates the presence of dark energy. This conclusion was largely determined by the object with largest $z$ observed to date. In 2001, a SN Ia with $z = 1.6$ was reported with larger apparent luminosity than expected in a matter dominated or in an empty Universe (Riess *et al.* 2001, see Figure 3). It can be reconciled with the small $z$ observations by assuming that it exploded in an epoch when the matter part of the energy density was still dominant and the rate of the expansion was decreasing. Quantitatively, the fit led to $\Omega_M = 0.35$, $\Omega_\Lambda = 0.65$.

The important question to be addressed in this context is the one regarding alternatives concerning the nature of the repulsive force. A cosmological constant implies an equation-of-state with $w$=-1 but any $w$<-1/3 yields repulsion. An additional, previously unknown, form of energy has been postulated as an alternative to the cosmological constant: **Quintessence** (Caldwell *et al.* 1998, Armendariz *et al.* 2000). It has repulsive properties but $w$≠-1. Therefore, it can be time-dependent and even have different values at different spatial points, contrary to a cosmological constant. Detailed SN Ia investigations try to put bounds not just on the size of the acceleration but also on the type of dark energy, i.e. the equation-of-state of the Universe.

The most recent analyses (Riess *et al.* 2007, Wood-Vasey *et al.* 2007) employing much larger data sets than before are all compatible with the cosmological constant ($w$=-1) interpretation of the data and give $\Omega_M = 0.274$ with 20% statistical error. A final conclusion concerning the acceleration driven by a substantial cosmological constant might be reached by the proposed Supernova Acceleration Probe space mission. The project aims at the observation of around 2000 SN Ia out to a redshift $z \leq 1.2$. Its launch is tentatively scheduled for 2013.

The results of 5 years of WMAP satellite mission were published and its cosmological interpretation was studied (Komatsu *et al.* 2009), taking into account the effect of the above listed investigations. The results were interpreted by assuming that our Universe is flat and its energy content is a mixture of ordinary matter, gravitating dark matter, and dark energy. The most important cosmological parameters were determined with unprecedented accuracy. The accuracy was substantially increased by combining the WMAP CMBR data, type Ia supernova luminosity distance measurements, and the largest scale components of the 2dF galaxy cluster catalogue (Percival *et al.* 2007). The supernova data are sensitive to the energy density component of cosmological constant type, while the galaxy clusters represent the aggregates of the gravitating (mainly dark) matter. This results in a value of the Hubble parameter $H$ at present time of $(70.1 \pm 1.3)\,\mathrm{km\,s^{-1}\,Mpc^{-1}}$. The ordinary matter content is $(4.62 \pm 0.15)\%$, the cold (non-relativistic) dark matter represents $(23.3 \pm 1.3)\%$, the part of the dark energy in the full energy density is $72.1 \pm 1.5\%$. The projection of the motion of such a universe back in time leads to a highly accurate estimate of its age: $13.73 \pm 0.12$ billion years.

Concluding, it becomes a more and more established fact that the chemical elements formed from baryonic matter contribute less than 5% to the total energy density of the present Universe. In view of the complete symmetry of laws governing matter and antimatter in our present day Universe it is actually a rather non-trivial fact that baryonic matter did not completely annihilate into radiation in the hot Universe directly after the inflationary epoch and that the original energy density rather was transformed into a high-temperature gas of ordinary elementary particles.

## 2.2   Origin of the matter-antimatter asymmetry

On the interface of neighboring domains of baryonic and antibaryonic matter, quark-antiquark (proton-antiproton) annihilation would lead to the emission of hard X-rays. The absence of this signal makes it highly probable that even if antibaryons were present in the early, hot Universe they had disappeared before the CMBR was emitted. Therefore, the observed baryonic density actually proves the presence of a matter-antimatter asymmetry within the present horizon in our Universe (Rubakov and Shaposhnikov 1996, Riotto and Trodden 1999).

In 1967 Sakharov analyzed the conditions which might lead to this asymmetry dynamically, instead of simply assuming it to be fixed by some initial conditions (Sakharov 1967). If, in a certain moment,

1. the elementary interactions violate the symmetry under the combined transformation consisting of spatial reflection (P) followed by charge conjugation (C) — the so-called CP symmetry,
2. the elementary interactions violate the baryon-antibaryon symmetry,
3. the Universe is out of thermal equilibrium,

then a matter-antimatter density difference is produced. One can deduce the actual amount of asymmetry with detailed quantitative calculations.

Without going into details, below we outline some of the scenarios proposed by theoretical particle physicists for the creation process of this fundamental asymmetry. For this we have to give a short account of the **Standard Model** of elementary interactions (Perkins 2000). (See also Chapter 8, Volume 1.)

Known elementary constituents of matter are quarks and leptons (see TABLE 1). Three families have been discovered. In each family one has two flavors of quarks and one lepton with the associated neutrino. The decay of the free neutron observed in 1932 and described first by the Fermi theory of weak interactions is understood today as the decay of a d-quark (one of three quarks composing the neutron) into a u-quark (which forms the final proton with the unchanged other two quarks) and an electron plus its antineutrino. The particles participating in this process constitute the lightest (1st) particle family of the Standard Model.

TABLE 1. Elementary particles in the Standard Model. For each family, the first line in the center column refers to quark species of left- and right-handedness (L,R), the second line gives the same for the leptons, not participating in strong interaction processes. The field quanta mediating strong, electromagnetic, and weak interactions are listed in the last column. They are not specific to any particle family.

| Particle families | Matter constituents (each has its own anti-particle) | Interaction vector particles |
|---|---|---|
| 1st | $(u_L, d_L)$, $u_R$, $d_R$ | |
| | $(e_L, \nu_{eL})$, $e_R$ | 8 gluons |
| 2nd | $(c_L, s_L)$, $c_R$, $s_R$ | photon ($\gamma$) |
| | $(\mu_L, \nu_{\mu L})$, $\mu_R$ | weak quanta ($Z^0$, $W^\pm$) |
| 3rd | $(t_L, b_L)$, $t_R$, $b_R$ | |
| | $(\tau_L, \nu_{\tau L})$, $\tau_R$ | |

Three elementary interactions act among these particles. Each of them is mediated by vector particles. The electromagnetic quanta, the photons, bind nuclei and electrons into atoms and molecules. Weak interactions are mediated by three vector fields, the $W^\pm$ and the $Z^0$, all discovered in 1983. Gluon fields bind the quarks into protons and neutrons. The strong interaction quanta come in eight different, so-called colored states and also each quark can appear in one of three different colored states.

It has been shown that the three Sakharov conditions might be fulfilled simultaneously in the Standard Model at high temperatures. The CP-violation, which allows the oscillation of the $K^0$ and of the $B^0$ mesons into their antiparticles and back, has been observed experimentally (in 1967 and 2001, respectively) and can be quantitatively understood with the present theory (Amsler *et al.* 2008).

On the other hand, no sign of baryon number violation has been observed to date in any elementary particle physics experiment. In the Standard Model one cannot find any process which would involve the transformation of a proton into mesons or leptons. However, in the early 1970's, G.`t Hooft (1976) showed that in the presence of specific configurations of electro-weak vector fields, fermions (leptons and quarks) can be created or annihilated, just the difference of the baryon number and of lepton number (B-L) should stay constant (quarks and antiquarks actually carry $\pm 1/3$ unit of baryon charge, while the lepton charge of the known species is $\pm 1$.). Today the chance for such transitions to occur is negligible (its probability is estimated to about $10^{-170}$). However they must have occurred frequently when the temperature was of the order of 100 GeV (about $10^{15}$ K).

It is a very interesting coincidence that exactly at that temperature scale one expects the transformation of all elementary particles from massless quanta into the massive objects observed in today's experiments. The creation of the mass is due to the so-called **Higgs effect**. This consists of the condensation of an elementary scalar field (a close relativistic analogue of the Cooper-pairing in superconductivity). If this transformation had proceeded via a sufficiently strong first order phase transition, the third of Sakharov's criteria had been also fulfilled by the behavior of the known elementary interactions in the very early Universe.

In a first order phase transition, the low-temperature (massive) phase would appear via thermal nucleation, which is a truly far-from-equilibrium process. Inside the bubbles of the new phase the baryon number-violating processes are stopped. So the question is this: What is the net baryon concentration frozen?

Quarks and antiquarks traverse the phase boundaries, which represent a potential barrier for them. As a consequence of the complex CP-violating phase in the Hamiltonian describing weak interactions, the reflection and transmission amplitudes for matter and antimatter turn out to be different leading to an asymmetry in the constitution of matter and antimatter inside the bubbles.

The quantitative details of this beautiful scenario critically depend on a single parameter: the strength of the self-coupling of the so called **Higgs field**, whose condensation determines the masses of all particles. This parameter is still unknown. The latest lower bound (Amsler *et al.* 2008) lies in a region where the phase transformation is actually continuous (beyond the end point of the first order transition line). The situation could be different in supersymmetric extensions of the Standard Model. The one explored best is the electroweak phase transition within the so-called Minimal Supersymmetric Standard Model (Carena *et al.,* 2009). One expects considerable guidance from measurements at the Large Hadron Collider in constraining the parameter space to search for the origin of baryon-antibaryon asymmetry. Another avenue could be the very late (low-energy density) exit from the inflationary period of evolution (Garcia-Bellido *et al.* 1999, Krauss and Trodden 1999, van Tent *et al.* 2004). If this energy scale coincides with the electroweak mass scale then the reheating of the Universe from its cold inflationary state would offer an out-of-equilibrium situation. This is the basis of the proposition of the Cold Baryogenesis scenario (Tranberg *et al,* 2007).

The resolution of the matter-antimatter asymmetry problem is an issue of central importance in particle physics in the twenty-first century.

## 2.3   Evolution of the expanding Universe

The equilibrium in the hot particle 'soup' is maintained through frequent elementary-particle reactions mediated by the quanta of the three fundamental interactions. The expansion of the Universe dilutes the densities and, consequently, the reaction rates get gradually lower. The adiabatic expansion lowers monotonically also the temperature (the average energy density). (Actually, there is a one-to-one mapping between time and temperature.) The following milestones can be listed in the thermal history of the Universe (Kolb and Turner 1990).

First, the weak interaction quanta became massive at the temperature scale of 100 GeV. Since then weak reactions have only occurred in contact interactions of the particles. At about the same time the t-quark and the Higgs quanta also decoupled from the 'soup'. The same decoupling happened for the other heavy quark species (b-quark, c-quark) and for the heaviest of the leptons (τ-particle) in the range 1-5 GeV (a few times $10^{13}$ K) of the average energy density. The τ-neutrinos remain in thermal equilibrium via weak neutral interactions.

The strong interaction quanta, the gluons became extremely short ranged at around the temperature $kT \sim$ 100-200 MeV (a few times $10^{12}$ K). Computer aided quantum field theoretical investigations have demonstrated that quarks and gluons are confined to the interior of nucleons (protons and neutrons) and excited baryonic resonances below this temperature range (Petreczky 2007). This transformation was smooth for baryonic matter densities characteristic of our Universe at that epoch and for the actual mass values of the light quarks, very similar to the process of atomic recombination.

At this stage no nuclear composite objects can be formed yet, since they would instantly disintegrate in collisions with hard electromagnetic quanta. The stabilization occurs for temperatures below 0.1 MeV. Primordial synthesis of light nuclei took place at that cooling stage of the Universe (Section 3).

Below this temperature light nuclei and electrons form a globally neutral plasma, in which thermal equilibrium is maintained exclusively by electromagnetic interactions. The gravitational attraction of the massive and electrically screened constituents of matter was balanced by the radiation pressure. This dynamical equilibrium is described by coupled fluid equations and results in acoustic oscillations modulating the essentially homogeneous distribution of the constituents (Hu 2001). The dominant wavelength of these oscillations is determined by those density fluctuation modes which left the horizon during inflation and are continuously reentering, since during the radiation dominated period the horizon expands faster than the global scale parameter of the Universe increases.

The last qualitative change occurred at the energy scale around 1 eV ($\sim 10^4$ K), when at the end of atomic recombinations the Universe became transparent to the propagation of electromagnetic radiation. At this moment the size of the Universe was about 1/1000 of its present radial scale. Today, the light emitted in the act of the last scattering is detected as Cosmic Microwave Background Radiation. A consistent interpretation of the details of its features represents (together with the primordial abundance of light nuclei) a unique test of all ideas concerning the earlier evolution of the Universe.

## 2.4 Gravitational clustering of matter

At the moment of the decoupling of light, the matter in the Universe became gravitationally unstable against density fluctuations. The key feature in understanding the emergence of a large-scale structure in the Universe is the statistical characterization of the density fluctuations at this moment. These fluctuations are determined by the spectra of the acoustic oscillations which are in turn determined by the reentering density fluctuations of inflationary origin. This line of thought leads us to the hypothesis of the quantum origin of the largest-scale structures observed in the Universe.

Amplitudes of density fluctuations at different wavelengths follow independent Gaussian (also called normal) statistics (see Subsection 3.6, Chapter 7, Volume 1), and their mean spectral power is distributed in an almost scale-invariant manner, described above. The absolute normalization was determined by the COBE satellite to be 1 part in 100 000. Their evolution can be analyzed initially with the help of the linearized gravitational equations. The classical analysis, originally performed by Jeans (1902), leads to the conclusion that fluctuations above the Jeans-scale are unstable and they are at the origin of the formation of the oldest structures (for a modern textbook on the subject, see Peacock 1999).

The non-linear stage of the clustering process can only be followed by numerical integration of Newton's equations of motion for a very large number (typically $10^6$ - $10^7$) of equal-mass particles. The most interesting question studied in these $N$-body simulations concerns the mass distribution of the first galaxies. It is this feature which determines the frequency of the occurrence of densities sufficiently high to start nuclear fusion reactions in these first gravitationally bound galactic objects.

The semi-empirical theory (Press and Schechter 1974) assumes that this distribution is determined by the probability of matter fluctuations obeying a Gaussian distribution to exceed an empirically determined threshold value. This simple idea results in an $\sim M^{-2/3}$ power scaling for the statistics of the collapsed objects with different mass $M$. This means that the earliest collapsed gas clouds were small, about $10^5$ solar masses, and had a temperature of a few hundred K. The thermal excitation of $H_2$ molecules provides the microscopic mechanism for the further radiative cooling, which might have led to the formation of the first minigalaxies and/or quasars. If sufficient quantities of $H_2$ molecules were present then the first stars and black holes were born very early, at a redshift $z \sim 20$ *i.*e. some 12 billion years ago, when the characteristic size of the Universe was $z+1$ times smaller than it is at present). If the dominant radiative cooling mechanism was the excitation of the atomic hydrogen, then the first objects with nearly $10^8$ solar masses were formed at a temperature of $10^4$ K. The first galaxies appeared in this case only for a redshift $z \sim 10$. One expects that astronomical surveys of the coming years will be able to reach the distance which can be calculated from the Hubble law $d = H/v$, where the Hubble parameter $H$ was defined in Subsection 2.1.4. The exploration of this distance scale should bring evidence for the existence of the first galaxies. According to the latest WMAP results (Komatsu *et al.* 2009) the first galaxies were formed at $z=10.8 \pm 1.4$.

The **Millenium Simulation** (Springel *et al.* 2005) is an N-body simulation tracing over 10 billion mass points, representing fractions of the primordial gas, from the time of the CMBR decoupling to the present-day Universe. This simulation showed that it was necessary to assume cold dark matter (consisting of slowly moving, heavy particles) to reproduce the large-scale structures found in galactic surveys. It was also able to show that bright quasars are formed already at very early stages, thus confirming observational results from the Sloane Digital Sky Survey (Anderson *et al.* 2001, Abazajian *et al.* 2009) which challenged traditional models of structure formation.

The radiative cooling and further gravitational evolution of the collapsed clouds leads to the appearance of the first stars (see Subsection 4.1.1).

# 3    PRIMORDIAL NUCLEOSYNTHESIS

## 3.1    Weak decoupling

After the quark-hadron phase transition no free quarks or gluons exist anymore. The hot plasma is composed of neutrons, protons, electrons, positrons, photons, and the electron-, muon-, and tau-neutrinos and their antineutrinos. Basically, all particles with masses $2m < kT/c^2$ are present because the respective particle-antiparticle pairs can be created in photon collisions. Scattering reactions thermalize all plasma constituents to the same temperature and forward and reverse reactions are in equilibrium. For instance, protons can be converted into neutrons by electron capture, $e^- + p \leftrightarrow n + \nu_e$, neutrons into protons by positron capture $e^+ + n \leftrightarrow p + \nu_e$. While all other constituents are highly relativistic, the nucleons are slower due to their large mass. Their kinetic energy exhibits a Maxwell-Boltzmann distribution appropriate for the given plasma temperature. Although formation and destruction reactions are in equilibrium, protons will be more abundant because of their lower mass. The ratio of neutron to proton number densities only depends on the temperature and the mass difference $\Delta m$ between the nucleons as long as the reaction equilibria apply (Börner 1988, Kolb and Turner 1990, Peacock 1999):

$$\frac{n_n}{n_p} = \exp\left(-\frac{\Delta mc^2}{kT}\right). \qquad (1)$$

Once $kT \leq 1$ MeV (i.e. at about $10^{10}$ K), the electrons are not energetic enough anymore to overcome the mass difference between neutrons and protons in electron captures. Also, photons cannot produce positrons anymore in pair-production processes to support positron capture on neutrons. Such weak transitions will thus cease to exist. Since the neutrinos were produced in such reactions, thermal communication of the neutrinos with the other constituents comes to an end. This phase is called weak freeze-out and weak decoupling because the neutrinos become decoupled from the rest of the particles and can assume different temperatures. Since there is little, if any, interaction between this neutrino background and the remaining particles, its evolution is governed predominantly by the expansion rate of the Universe similar to the photon background radiation after electromagnetic decoupling that gives rise to the cosmic microwave background (see Subsection 2.1.1). Thus, in addition to the cosmic microwave background there is a cosmic neutrino background stemming from the era of weak decoupling. Its temperature is lower by a factor of 0.714 because of the heating of the photons by $e^+$ - $e^-$ annihilation after weak decoupling.

With the ceasing of the electron and positron captures, the ratio of neutrons to protons gets frozen at the decoupling temperature, yielding a value of about 1/6. However, after the weak freeze-out, photons still dominate the total energy of the Universe and thus the temperature is decreasing as the inverse square root of the time, according to the law valid for a radiation-dominated Universe. The ratio between baryon and photon number densities $\eta = n_b / n_\gamma$ characterizes a particular solution of the equations for the expanding early Universe and, therefore, the solutions can be labeled by the parameter $\eta$ (Kolb and Turner 1990, Coles and Lucchin 1996, Riotto and Trodden 1999). Assuming a globally valid $\eta$, the baryon density can be written as a function of temperature: $\rho_b = 3.376 \times 10^4 \eta\, T_9^3$ g cm$^{-3}$ ($T_9$ is the temperature in units of $10^9$ K.) In fact, $\eta$ is inversely proportional to the total entropy of the Universe, which has to remain constant.

Together with the equation for the evolution of the temperature $T_9 = 13.336 / t^{1/2}$, this sets the conditions for primordial nucleosynthesis. The strength of the standard big bang scenario is that only one free parameter—the above introduced baryon-to-photon ratio $\eta$—must be specified to determine all of the primordial abundances ranging over 10 orders of magnitude.

The parameter $\eta$ also depends on $\Omega_b = \rho_b/\rho_c$, i.e., the ratio of the baryon density to the critical density $\rho_c$ needed for a flat Universe.

Thus, a fit of $\eta$ to observed primordial abundances not only probes the conditions in the early Universe at the time of nucleosynthesis but can also reveal the curvature of the Universe, or at least the baryonic contribution to that curvature (Schramm and Turner 1998). Historically, primordial nucleosynthesis was the first tool for determining the geometry of the Universe. With the increased accuracy in the resolution of the angular multipole expansion of the CMBR temperature fluctuations delivered by WMAP, the total density of the Universe (and not just the baryonic one) can be determined independently now (see Section 2.1).

## 3.2 The reaction network and the production process of nucleosynthesis

After the weak freeze-out, the baryonic matter part essentially consists of free neutrons and protons interacting with each other. Deuterium is constantly formed via neutron captures on protons. However, due to the low binding energy of the deuteron the created deuterons will preferably be destroyed by photodisintegration as long as the (photon) temperature is higher than $10^9$ K. Below that temperature, photodisintegration is no longer effective, and more heavy elements can be built up by further reactions on the deuterons. Thus, although free neutrons and protons had already existed earlier, the onset of further primordial nucleosynthesis is delayed until about 2 s after the Big Bang (Boesgaard and Steigman 1985, Bernstein *et al.* 1991).

Because free neutrons are not stable, but decay with a half-life of $T_{1/2} = (10.25 \pm 0.015)$ min, the neutron-to-proton ratio will change from 1/6 to 1/7 until the onset of primordial nucleosynthesis.

While the Universe expands further, it cools down and the decreased energy of the photons cannot prevent significant formation of deuterons anymore. Thereafter, heavier nuclides can be created by reactions involving protons, neutrons, and the newly formed nuclear species (Schramm and Turner 1998, Sarkar 1996). This is nothing else than a freeze-out from a high-temperature, low-density nuclear statistical equilibrium (NSE), similar to the one occurring in late and explosive phases of nucleosynthesis (see Sections 4.4 and 4.5) but at different density. An NSE is defined by all reactions via the strong and electromagnetic forces being in equilibrium. The equilibrium abundances are then only determined by the (baryon) density, the temperature, and the binding energy of the nuclei, as well as by the initial composition of the material, i.e. the neutron-to-proton ratio. The latter is set by the weak freeze-out and the subsequent neutron decay. In a high-temperature NSE, all nuclei are completely dismantled into their constituents, the protons and neutrons. Assuming a quick freeze-out in such a manner that late-time non-equilibrium reactions will not significantly alter the NSE abundances, the resulting abundances can already be determined without detailed reaction network calculations. With decreasing temperature, simply the most strongly bound nuclei at the given neutron-to-proton ratio will be formed. Thus, we expect to find mainly unprocessed protons and $^4$He nuclei, exhibiting a high binding energy, with all neutrons having been incorporated into the $^4$He nuclides. Formation of elements beyond He is hindered by the fact that there are no nuclei with mass numbers 5 and 8. The $3\alpha$ reaction (see Section 4.3) could convert $^4$He to $^{12}$C but is not in equilibrium because it is strongly dependent on the density and too slow at the conditions in the early Universe. Nuclei close to $^4$He are produced according to their binding energies.

The important reactions and the produced nuclear species are shown in Figure FIGURE 4. The conditions at the onset of and during the nucleosynthesis are given by the initial values and parameters discussed above, with $\eta$ being a free parameter.

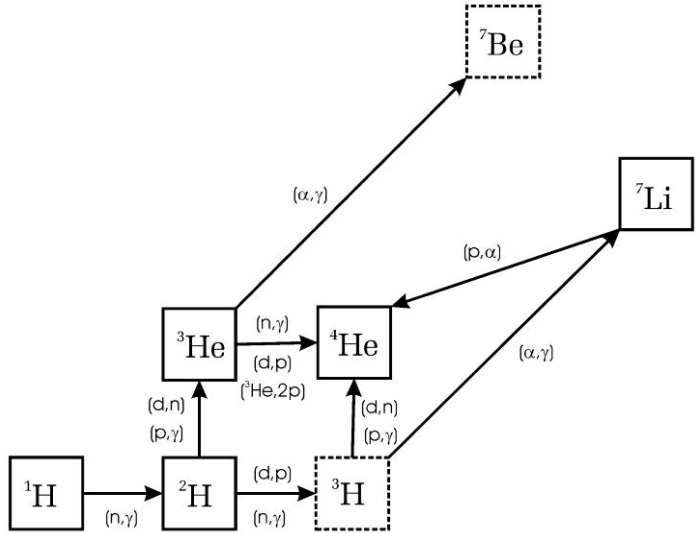

FIGURE 4. The reaction network of standard big bang nucleosynthesis. Unstable nuclear species are marked by dashed boxes. When all reactions are stopped, the unstable $^7$Be decays to $^7$Li and $^3$H decays to $^3$He.

A typical result of a full network calculation for a specific $\eta$ is shown in Figure 5 (Tytler *et al.* 2000, Kolb and Turner 1990). As can be seen, practically no nucleosynthesis occurs during the first two seconds the temperature remains above $10^{10}$ K and no other nuclides than free nucleons are favored. Only after a sufficient drop in temperature deuterons are formed, $^3$H and $^3$He nuclei are produced, quickly followed by $^4$He. The neutron abundance is determined by slow beta-decay in the early phase. During the formation of $^4$He the neutron abundance suddenly drops because most of the neutrons are incorporated into the $\alpha$ particles. Slightly delayed, traces of $^7$Li and $^7$Be nuclei are formed. During the decline of the temperature, charged particle reactions freeze out quickly and after about 15-30 minutes nucleosynthesis ceases. The very few free neutrons decay into protons, $^3$H eventually also decays and so does $^7$Be which forms further $^7$Li. Finally, the by far most abundant species are hydrogen (protons) and helium ($^4$He, i.e. $\alpha$ particles) which together give more than 99.9% of the baryonic material. Although this calculation did not assume NSE at all times, the resulting abundances are very close to the ones obtained from equilibrium abundances and fast freeze-out. Realizing the dominance of $^4$He due to its high binding energy, it is easy to understand that about 25% of the gas are made of helium. The initial n/p ratio of 1/7 translates into mass fractions $X_n = 0.125$ and $X_p = 0.875$ (i.e. 12.5% of the gas mass consists of neutrons). Assuming that all neutrons combine with protons to form $^4$He, the mass fraction of $^4$He has to be $X_\alpha = 2X_n$ because it contains 2 neutrons and 2 protons. Thus, $X_\alpha = 0.25$, i.e. 25%.

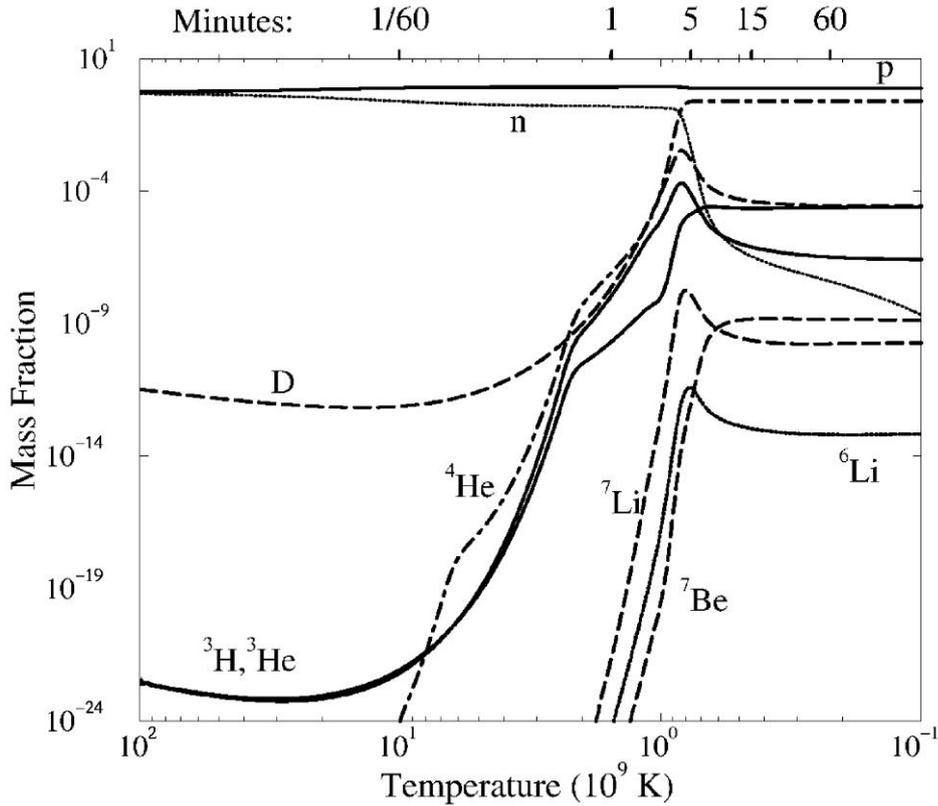

FIGURE 5. Primordial abundances of different nuclear species as a function of time and temperature for a fixed ratio of baryon to proton number densities, $\eta = 5.1 \times 10^{-10}$. [Reprinted from Tytler *et al*. 2000 with kind permission of the first author and IOP.]

An interesting result was obtained from the fact that the initial n/p ratio depends on the weak freeze-out (see above). The freeze-out time and temperature and thus also the resulting n/p ratio depends on the change in the degrees of freedom during freeze-out (Kolb and Turner 1990). Although only three neutrino families were known originally ( $N_\nu = 3$ ), it remained an open question whether there are more families containing light (nearly massless) neutrinos. Increasing the number of neutrino species increases the number of degrees of freedom and has a similar effect to that of a faster expansion resulting from larger pressure. Thus, a larger number of neutrino families would lead to an earlier weak decoupling at a still higher temperature. This results in a higher n/p ratio and consequently leads to more $^4$He.

Primordial nucleosynthesis calculations performed along these lines were able to put the limit $N_v \leq 3.3$, ruling out any further families long before any particle physics experiments. This was possible due to several facts: helium is very abundant and can be easily measured in astronomical observations, it is strongly bound and therefore robust during galactical chemical evolution, its abundance only weakly depends on $\eta$, the primordial production reactions are well determined, and the half-life of the neutron (determining the change of the n/p ratio between weak freeze-out and onset of nucleosynthesis) is measured to high accuracy. This prediction on the number of neutrino species was later confirmed by experiments at CERN (see the article of D. Karlen in Hagiwara *et al.* 2002) directly measuring the decay widths of the $Z^0$ boson in the weak interaction.

The nuclear reaction rates (cross section) for the reactions specified in Figure FIGURE 4 are well determined, also at the interaction energies relevant to primordial nucleosynthesis which are comparatively low by nuclear physics standards. Thus, once the initial conditions are determined, the evolution of the different species with time and the final abundances can be calculated with high accuracy. The only open parameter in the standard Big Bang nucleosynthesis model is $\eta$. Since the baryon density is proportional to $\eta$ and the reaction rates are density dependent, the final abundances will also depend on the choice of $\eta$.

Figure 6 shows this dependence for a typical range of the baryon density and equivalently its fraction of the critical density $\Omega_b$. Immediately catching the eye is the difference between the density dependences of the different species: the $^4$He fraction increases slightly with increasing density while the deuterium and $^3$He fractions strongly decrease (note that the vertical scale of the upmost curve is linear, while those of the rest are logarithmic). The simple explanation is that a high density during the nucleosynthesis phase gives rise to a larger number of capture reactions on deuterium and $^3$He, producing more $^4$He but leaving less of the targets. $^7$Li shows a more complex behavior with a pronounced minimum. At low densities $^7$Li is produced by $\alpha$-capture on $^3$H but is preferably destroyed at higher densities by the reaction $^7$Li(p,$\alpha$)$^4$He. However, at higher densities the production of $^7$Be via $^3$He($\alpha,\gamma$)$^7$Be also increases. This $^7$Be is not destroyed by any further primordial process but eventually decays to $^7$Li after nucleosynthesis has ceased. The different density dependences of the three main reactions involved give rise to the minimum shown in Figure 6.

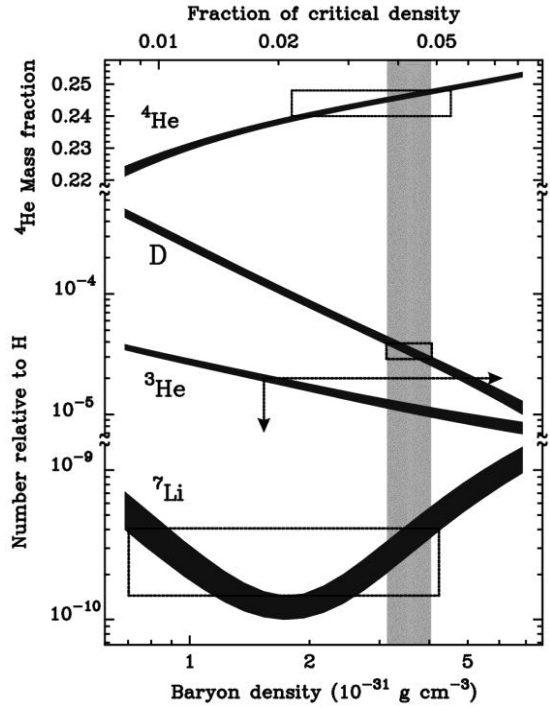

FIGURE 6. Primordial abundances as a function of baryon density $\rho_b$ or fraction of critical density $\Omega_b$ (these two quantities are directly related to the baryon-to-photon ratio $\eta$, see text). The widths of the curves give the nuclear physics uncertainties. The boxes specify the ranges of abundances and densities constrained by observation (there is only an upper limit for $^3$He from observation) as given in Burles *et al.* 1999, 2001. $^4$He is very abundant and thus it can be observed with high accuracy. The shaded area marks the density range consistent with all observations (Burles *et al.* 1999, 2001). [Reprinted from Tytler *et al.* 2000 with kind permission of the first author and IOP.]

## 3.3 Comparison of calculations and observed primordial abundances

Except for helium, the abundances of the primordial isotopes change by orders of magnitude when varying the baryon density or $\eta$, respectively. On the other hand, $^4$He is very abundant and thus it can be observed with high accuracy. Therefore, it is possible to determine $\eta$ from comparison with primordial abundances. As outlined above, this means nothing less than determining the total baryon density and thus the baryonic contribution to the curvature of our Universe! The allowed abundance range for each primordial species is shown in Figure 6, combined with the calculated baryon density. It is reassuring that the independent observations for each species still permit a consistent range of $\eta$, i.e., all the observations are consistent with the same baryon density.

Serious complications arise from the fact that the calculations have to be compared to primordial abundances, not to the ones currently found in stars or the interstellar medium (Boesgaard and Steigman 1985, Olive *et al.* 2000, Tytler *et al.* 2000, Olive 2001, Steigman 2007). During the further evolution of the Universe, when stars are formed and destroyed, the abundances within a galaxy change with time (see Subsection 6.4) due to the stellar production and destruction processes (see Section 4). Thus, it is not trivial to determine the primordial values, especially those for isotopes which are weakly bound and therefore easily destroyed in stars such as deuterium, $^3$He or $^7$Li nuclei. In order to trace the effects of chemical evolution, the standard practice is to correlate the abundance of the nuclide in question with the abundance of a nuclide which becomes monotonically enriched in the interstellar medium during the course of the chemical evolution, e.g. oxygen. This other nuclide serves as a tracer of the metallicity of an object, i.e. the content of elements beyond hydrogen and helium (a technical term borrowed from astronomy). From the observation of different sites with different metallicities, one hopes to find trends which can be extrapolated to zero metallicity, i.e. to the primordial values. Figure 7 shows an example of such an extrapolation.

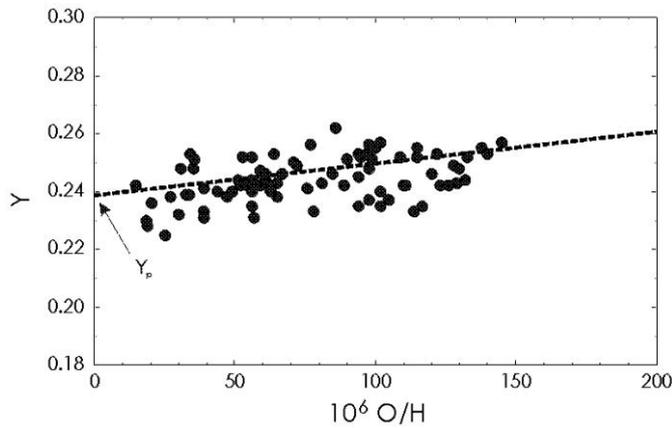

FIGURE 7. Example for the extrapolation to primordial values. The primordial value of $^4$He ($Y_p$) for zero metallicity (zero oxygen) is inferred from a series of observations (filled circles) (Olive 2001). For other primordial isotopes the slope of the interpolation would be steeper, for $^7$Li there would be a steep rise at higher metallicities and a flat line with nearly constant abundance at low metallicity. [Reprinted from Olive (2001) with permission by Springer.]

The spectra of $^4$He are easy to observe due to its high abundance. This He isotope can be identified in clouds of ionized interstellar gas around bright, young stars (HII regions). Although the abundance in an object can be determined with high precision, a large systematic uncertainty remains due to observations of different objects. On the other hand, the observation of $^3$He is difficult and an extrapolation carries a large uncertainty because different stars both produce and destroy it effectively. Due to its weak lines it cannot be seen in low-metallicity stars. The chemical evolution of $^7$Li is also complex but it can at least be seen in old stars. Such stars contain a factor of 0.03-0.0003 less 'metals' (in astronomy: all elements except H and He) than the Sun. Studying such stars, a remarkable discovery was made: below a certain metallicity the content of $^7$Li remains almost constant, i.e. a plateau appears in the abundance plots. The $^7$Li-plateau was thought to give the primordial value. However, the question of possible depletion in early stars remains to be addressed (Korn *et al.* 2006) as well as the dependence of the data on the underlying assumptions, such as model atmospheres (Charbonnel and Primas 2005; Asplund *et al.* 2006a). These problems give rise to a large systematic uncertainty also for $^7$Li. Until recently deuterium was usually observed in the local interstellar medium. However, the situation has greatly improved in recent years. It became possible to observe D absorption lines in quasar spectra (Tytler *et al.* 2000). Quasar light passes through clouds at high redshift, i.e. far away and at the same time far back in time, which contain (almost) primordial amounts of D. Because of the brightness of quasars such lines can still be identified although the photons have traveled a long distance. In summary, only upper limits can be given for primordial $^3$He and the primordial values of $^4$He and $^7$Li have a smaller individual observational error compared to $^3$He but a considerable systematic uncertainty in the observations. The averaged D values from far objects seem to give the tightest constraint for Big Bang Nucleosynthesis (BBN) calculations (Steigman 2007).

The dependence of primordial abundances on $\Omega_b$ are shown in Figure 6. Using the available abundance information, it is found that it has to lie between about 0.038 and 0.048, also somewhat depending on the choice of the value of the Hubble constant (Schramm and Turner 1998, Burles *et al.* 2001, Steigman 2007). Most recently, an independent value for $\eta_b$ was derived from the WMAP data of the CMBR (see Komatsu *et al.* 2009 for latest results). The CMBR value of $\eta_b$=(6.11±0.2)×10$^{-10}$ is in excellent agreement with the value derived from the BBN abundances for D and $^3$He. Depending on improvements in the observations and theoretical models of the evolution of $^4$He and $^7$Li with metallicity, their required $\eta_b$ may or may not remain in agreement with the CMBR value. However, the BBN model still is in good standing when considering the CMBR data due to the large systematic uncertainties inherent in the observations of the latter isotopes (Steigman 2007).

Assuming only baryonic matter, the small value of $\Omega=\Omega_b$ would indicate an open Universe, i.e., one expanding forever. However, theory of inflation demands that $\Omega = 1$ (exactly) and thus a flat Universe (Kolb and Turner 1990, Peacock 1999). This is in agreement with the latest observations of the CMBR as discussed in Subsection 2.1.2. Then the missing mass required to close the Universe must be non-baryonic. Indeed, there are other indications (in the initial creation and later dynamics of galaxies) that there is more gravitational interaction than can be accounted for by standard baryonic matter. Of further interest is the value of the cosmological constant (see Subsection 2.1.4), which was found to be non-zero in recent distance measurements (Perlmutter *et al.* 1997, Schmidt *et al.* 1998, Perlmutter *et al.* 1999, Wood-Vasey *et al.*

2007, Riess *et al.* 2007), using type Ia supernovae as 'standard candles' (Subsections 2.1.4, 5.3). The cosmological constant $\Lambda$ also provides a contribution $\Omega_\Lambda$ to the expansion of the Universe so that the new requirement reads $\Omega_M + \Omega_\Lambda = \Omega_b + \Omega_s + \Omega_\Lambda = 1$, with $\Omega_s$ being the non-baryonic contribution (Steigman *et al.* 2000). At the time of writing, it is not clear yet what constitutes the non-baryonic mass. A quite general name for possible new, exotic particles is WIMP, an acronym for 'weakly interacting massive particle' (see Sections 2.1.3 and 2.1.4 for details).

# 4    STELLAR NUCLEOSYNTHESIS

## 4.1    Stellar evolution

### 4.1.1  Birth of stars

Stars are formed within molecular clouds, vast aggregations of molecules residing in the galactic disks. These clouds which often contain the mass of a million stars, are much denser and colder than the surrounding interstellar gas. Stars are born out of the collapse of small condensation areas that are scattered throughout the much larger volume of a molecular cloud. The collapse can occur due to random density fluctuations or be externally triggered, e.g., by shockwaves from supernovae or galaxy collisions. Soon after the collapse begins, a small pressure-supported protostar at the very center of the collapse flow develops. During the main collapse phase, the central protostar is surrounded by an inward flow of gas and dust. As the protostar evolves both the temperature and the density increase inside. Finally, the central core of the protostar heats up so much that nuclear 'burning' is initiated and the star begins its energy production through nuclear fusion.

Star formation is a process complicated by the details of cloud fractionation, rotation, turbulence, and magnetic fields. While the formation of low mass stars (below 8 solar masses) is thought to be understood and proceeding through an accretion disk, the mechanism to form more massive stars is not understood as well. Due to the larger radiation pressure of their emissions, the accretion disk would be blown away. The current model assumes, consistent with observations, the formation of a directed jet, transporting a small fraction of material but clearing a cavity through which most of the radiation can escape without interaction with the accretion disk (Bannerjee and Pudritz 2007). In this way, low mass and high mass stars could be formed in a similar manner. Other models assume coalescence of two or more light stars or competitive accretion of a low and a high mass star feeding from the same molecular cloud (Bonnel, et al., 1997; Bonnel and Bate 2006).

The galactic mass distribution of the newborn stars is known as the initial mass function. To sustain nuclear burning in their interiors stars must at least have 8 percent of the mass of the Sun. During their formation, stars with a smaller mass do not release sufficient gravitational binding energy to heat the gas to temperatures required for igniting nuclear fusion. These are called brown dwarfs. The lower-mass stars between about 8 and 40 percent of the mass of our Sun are called red dwarfs, because of their small size and their low surface temperature. At the other end of the mass range stars more than 100 times as massive as the Sun are highly unstable due to spontaneous pair production of electrons and positron from plasma interactions and therefore do not exist in our Universe. During their enormous life spans stars produce energy through nuclear fusion and shine continuously over millions to billions of years. Lower-mass stars consume their fuel very quietly and survive for several billion years. Massive stars, on the other hand, burn out in a few millions of years.

### *4.1.2  Hertzsprung-Russel diagram*

Stars undergo drastic changes during their evolution. One of the best methods for charting the course of stellar evolution is the Hertzsprung-Russell (HR) diagram shown in Figure 8, a particular type of graph developed in the early 20th century by the astronomers Hertzsprung and Russel. In this diagram the luminosity or energy output of a star is plotted on the vertical axis, and the surface temperature of the star on the horizontal axis. For historical reasons, the surface temperatures along the horizontal axis are plotted backwards, so that they increase toward the left. In the HR diagram the various stars are then plotted according to their luminosity and surface temperature. As one can see, the stars are not distributed randomly in the HR diagram, but are rather grouped in certain areas.

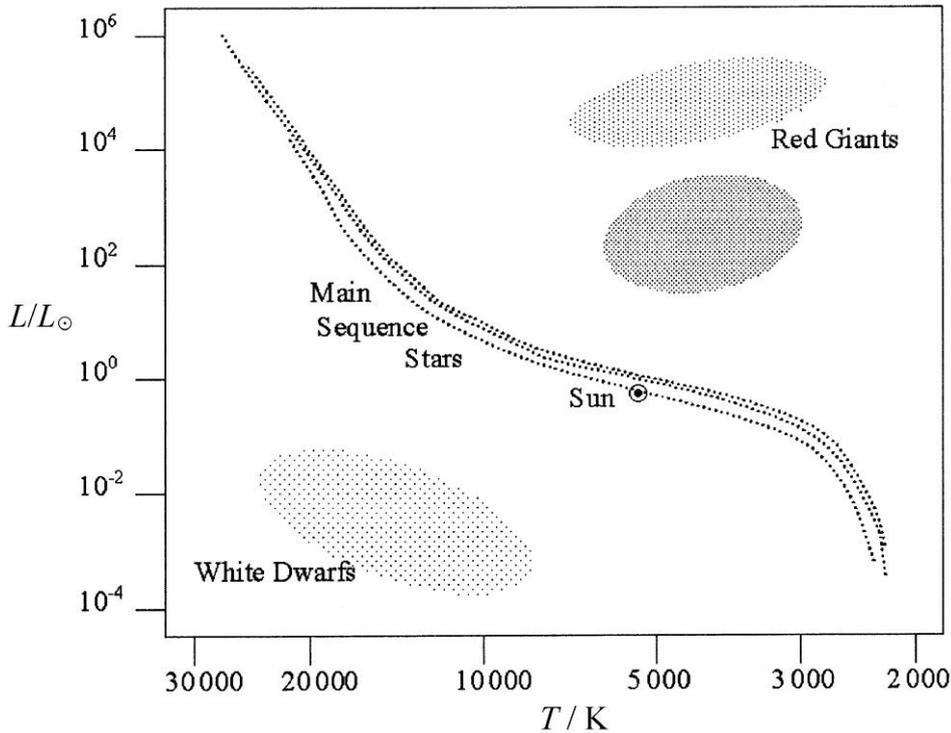

FIGURE 8. The Hertzsprung-Russell (HR) diagram. In this schematic diagram the various stars are plotted according to their relative luminosity $L/L_\odot$ (where $L_\odot$ is the absolute luminosity of the Sun) and their surface temperature, $T$.

Most of the stars line up along a well-defined band on the HR diagram known as the main sequence and are therefore also called main-sequence stars. This trend is no coincidence. Stars that lie along the main sequence have the proper internal configurations to support fusion of hydrogen to helium. Since stars spend most of their lifetime in this hydrogen burning state, most stars in the HR diagram are lying on the main-sequence band. Our Sun is also a typical main-sequence star.

After the hydrogen supply in the core of the star is exhausted and converted to helium, the central temperature is too low to fuse helium into heavier elements. Therefore, the core lacks an energy source and cannot support anymore the overlying bulk of the star. Through the gravitational pressure the size of the core shrinks and the temperature of the central region increases accordingly. The heat released by the core increases steadily the luminosity of the star. Paradoxically, even though the helium core is shrinking, the radius of the star, determined by the outer hydrogen layer, increases by factors of 100 to 1000. Through this expansion, the surface temperature is reduced up to 50 percent and the star becomes redder. Therefore, these stars are called red giants and are found in the HR diagram in the upper right corner.

When the core temperature of the red giant reaches about one hundred million degrees, a new sequence of nuclear reactions called helium burning begins in the core where helium nuclei fuse to carbon and oxygen. Our Sun has lived for 4.5 billion years and has already burnt half of its hydrogen in the core. After about another 5 billion years our Sun will also become a red giant and will thereby increase its size so much that the radius of the Sun will reach about the Earth's orbit.

The further evolution of a star and the nature of its stellar death depend on the initial mass. If the initial mass of a star is less than about 8 solar masses, it is burning He unstably (see also the section on He burning) and the resulting pulsations lead to the loss of huge quantities of hot gases towards the end of its life. This cloud moving away from the star is called a planetary nebula. The central small and hot core of the star that is left over is a white dwarf and consists of the ashes of helium burning, i.e., carbon and oxygen. Even though the surface temperature of the white dwarf is still very hot its luminosity is small, because nuclear fusion has ceased. Therefore, the white dwarfs are found in the lower left corner of the HR diagram.

If the initial mass of a star is more than about 8 solar masses further burning phases will take place. These are called advanced burning phases and consist of carbon, neon, oxygen, and silicon burning, being named after the nuclei mainly destroyed in that phase. In these subsequent burning phases heavier and heavier nuclei are built up, and the ashes of the preceding burning phases provide the fuel for the subsequent burning phases. However, in the outer and therefore cooler and less dense regions of the star the previous burning phases are still continuing. This leads to shell burning with distinct adjacent shells of different chemical compositions, in which different burning phases prevail. In the outermost shell of the star still hydrogen is burnt into helium (hydrogen burning), in the next shell helium to carbon and oxygen (helium burning), and finally, in the fully evolved star, there follow still carbon, oxygen, neon and silicon burning shells (Figure 9). In the core of the star significant amounts of iron are accumulating through silicon burning. A detailed discussion of the nuclear burning phases is given in Subsections 4.2 through 4.4.

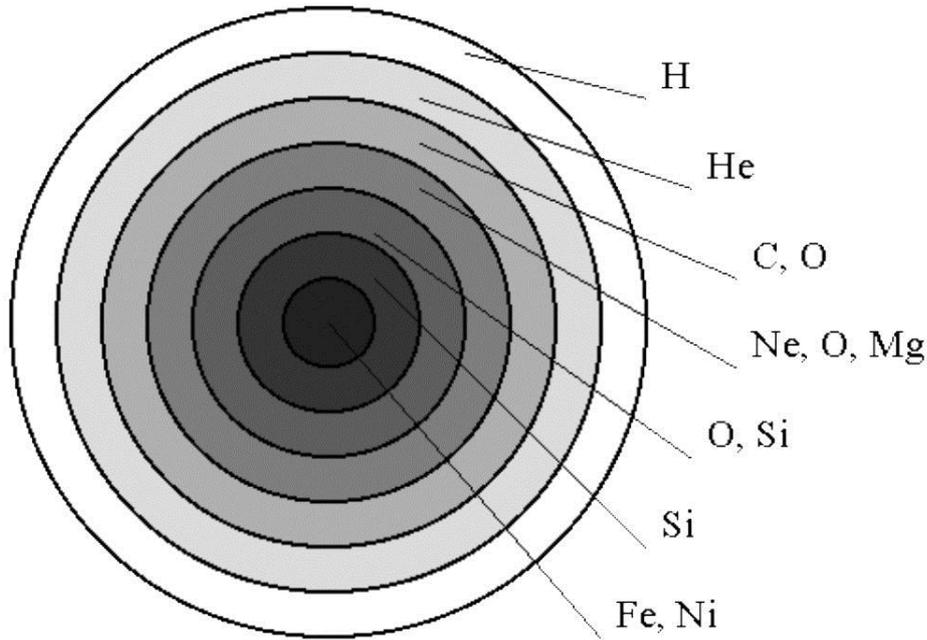

FIGURE 9. Schematic structure of the different shells of a fully evolved massive star with their prominent constituents. At the bottom of each shell, different burning phases take place. The iron core in the center has been accumulated in silicon burning. The outer part of the star consists of a thick envelope of unburned hydrogen. Note that the relative thickness of the layers is not drawn to scale.

A detailed introduction to stellar evolution is given in the following books: Clayton 1984, Hansen and Kawaler 1994, Kippenhahn and Weigert 1994, Phillips 1994, Tayler 1994. Books and reviews discussing stellar nucleosynthesis are Rolfs and Rodney 1988, Arnett 1996, Thielemann *et al.* 2001b. Tables of nuclear reaction rates and cross sections can be found in Rauscher and Thielemann 2000.

### *4.1.3 Supernova explosions*

The endpoint in the evolution of stars with more than 8 solar masses is a type II supernova. One should not confuse novae with supernovae and even the two types (i.e. type I and type II) of supernovae are quite different. It will become evident in the following that the sites of these explosive events are only loosely related, despite the similarity in the name. There is a major difference in the underlying mechanism between type I (SN I) and type II supernovae (SN II). The confusing choice of names is, once again, historical. Astronomy is guided by observations and early astronomers did not have the equipment to investigate the objects in any detail. Obviously, even today it is impossible to directly view the events in binary systems, but much more detail in light curves (i.e. brightness as a function of time) and spectra can be studied.

Historically, the comparatively frequent novae (see Subsection 5.2) were named first. Subsequently, much brighter eruptions of light were observed in the sky. Since they are brighter by more than a factor of $10^6$, they were appropriately termed supernovae. The light curve of a supernova is somewhat different from that of a nova: its rise time is only a few hours instead of days and it exponentially decays after having reached its peak. Closer investigations showed that several classes of supernovae can be found, according to features in their spectra: type I do not show hydrogen lines, whereas they are found in type II eruptions. This indicates whether the exploding object has an extended hydrogen envelope (such as massive stars). A more detailed classification scheme is shown in TABLE 2. Type Ia supernovae are further discussed in Subsection 5.3, while the unique scenario producing all other types (SNIb,c; SNII) is shortly introduced in the following.

TABLE 2. Supernova classification by observed properties. Different lines can be identified in the spectra. The assumed scenario causing the light burst is also given.

| Characteristic explosion energy ($10^{44}$ J) and light curve | | | | |
|---|---|---|---|---|
| no H lines | | | H lines | |
| Si lines | no Si lines | | Exponential decay of light curve | Plateau feature in light curve |
| | He lines | No He lines | | |
| White dwarf disruption | Core collapse (binary system?) | Core collapse (binary system?) | Core collapse | Core collapse |
| **SN Ia** | **SN Ib** | **SN Ic** | **SN II L** | **SN II P** |

### 4.1.4 Core collapse supernovae

When the stellar core becomes dominated by iron, the fusion into heavier elements does not lead to the release of energy, but rather requires absorption of energy. (Editors' note: See, e.g., Figure 13 in Chapter 5, Volume 1, showing the cross section of the stability valley of nuclei with iron at its lowest point.) Therefore, the core lacks an energy source and is unable to support itself against gravity anymore leading to a collapse of the star. In a single second the innermost regions are compressed to nuclear densities of about $10^{12}$ kg/m$^3$ and temperatures of about $10^{11}$ K. The iron nuclei, which have been synthesized just before in silicon burning are broken up again into protons and neutrons through the high-energy thermal radiation. The innermost regions are compressed so much that the core density becomes sufficiently high for electrons and protons to combine, producing neutrons and neutrinos. As the collapse continues, this giant ball of neutrons generally reaches a state of maximum density, and then bounces back. The bounce drives an extraordinarily powerful shock wave outwards through the outer parts of the star. Investigations within the last three decades have made it clear that this prompt shock will not have enough energy to explode the remaining outer layers of the star. Only with the additional supply of energy through neutrino heating can the shock wave be supported to completely blow apart the star. This powerful explosion can explain supernovae of type II, but also of type Ib,c. At the center of the supernova explosion, the dense core of neutrons may be left behind as a neutron star. Alternatively, if the remaining core becomes heavier than a few solar masses through partial fallback of material, it can even collapse into a black hole. The dual explosion mechanism with prompt shock and delayed explosion by neutrinos is still not understood well. A proper treatment of the neutrino transport requires detailed 3D hydrodynamical simulations which are currently beyond the capability of modern computers and thus one has to refrain to approximations whose merits are debatable. For an overview, see, e.g., (Janka, et al., 2007).

The physics of the remaining compact objects after stellar death, e.g., white dwarfs, neutron stars and black holes are discussed by Shapiro and Teukolsky (1983).

## 4.2  Hydrogen burning: proton-proton chain, CNO cycle

For nuclear reactions to take place in the interiors of stars at least a temperature of 10 million degrees is necessary. This high temperature is needed because nuclei are positively charged and repel each other through the Coulomb potential. The typical kinetic energy of nuclei in stellar interior range from between a few keV to a few 100 keV being much smaller than the typical height of a few MeV of the Coulomb barriers between reaction partners. Therefore, nuclear reactions in stars proceed mainly by barrier penetration exploiting the quantum mechanical tunnel effect. The cross sections decrease exponentially with the kinetic energies of the nuclei because of the decreasing penetration probability through the Coulomb barrier. The dependence on the relative kinetic energy $E$ between interacting nuclei can be represented most simply by a formula in which a factor proportional to the inverse of the relative kinetic energy $1/E$ and the barrier penetration factor $G(E)$ is factored out from the cross section: $\sigma(E) = (1/E) \, G(E) \, S(E)$. This leaves a function $S(E)$ called the astrophysical $S$-factor that varies smoothly with the kinetic energy $E$ of the interacting nuclei in the absence of resonances. Neutron-induced reactions would not have to overcome the Coulomb barrier. However, neutrons are not very abundant in stellar interiors. They still play a major role for the nucleosynthesis of heavy nuclei through the so-called s- and r-processes, to be discussed in Subsection 4.5.

The reaction rate, expressed as the number of reactions per volume and per time, is proportional to the astrophysical $S$-factor. At the temperatures and densities relevant for the stellar environments the interacting nuclei have a Maxwell distribution of speeds. This distribution has also to be taken into account when determining the reaction rate. An introduction to astrophysical $S$-factors and reaction rates can be found in many textbooks on nuclear astrophysics, e.g. Arnett 1996, Rolfs and Rodney 1988, Iliadis 2007, Boyd 2008.

Nuclear burning in late hydrostatic phases (see, e.g., silicon burning) and in different explosive scenarios proceeds at high temperatures and densities. This leads to equilibrium between forward and reverse reactions, e.g. capture and photodisintegration. It gives rise to equilibrium abundances depending only on the supply of free neutrons and protons and on certain nuclear properties. High temperatures favor the creation of light nuclei because the photodisintegration processes dominate. High densities lead to heavy nuclei, and intermediate conditions yield the highest abundances for nuclei with high binding energies. Such an equilibrium can be established within a group of nuclear species where individual reactions link different groups. This is called quasi-statistical equilibrium (QSE). The full nuclear statistical equilibrium (NSE) is reached when all nuclei are equilibrated.

In the following we describe the different burning stages one by one, starting with hydrogen burning, being the first burning stage of every star.

In hydrogen burning, occurring in the cores of main-sequence stars like our Sun, ordinary hydrogen nuclei (i.e. protons) are burnt through a chain or cycle of nuclear reactions into $^4$He nuclei. In this stellar plasma there are two processes burning hydrogen: the proton-proton chain (pp-chain) (Figure 10) and the CNO-cycle (Figure 11).

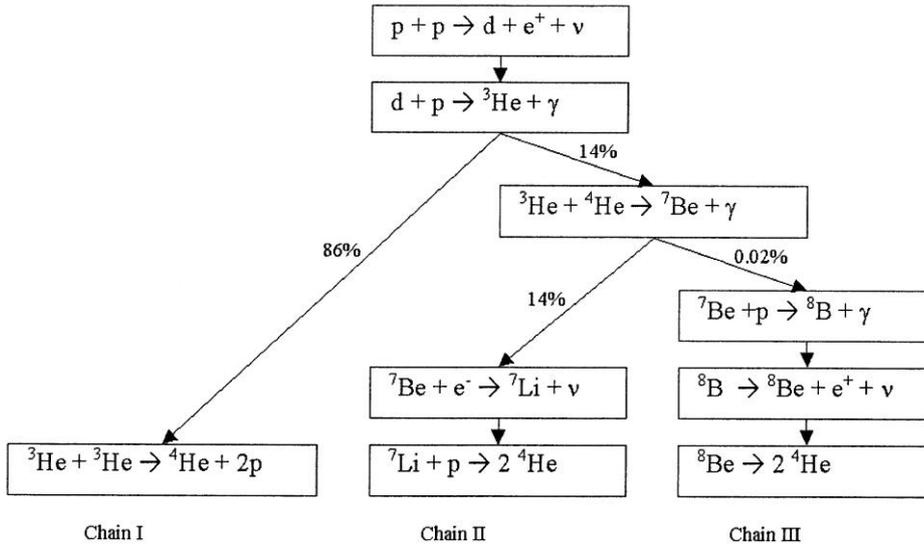

FIGURE 10. The proton-proton chain of hydrogen burning.

The pp-chain proceeds through a sequence of two-body reactions. The first reaction in the pp-chain is the exothermic fusion of two protons p into the deuteron d consisting of a proton p and a neutron n through the reaction:

$$p + p \rightarrow d + e^+ + \nu \,. \qquad (2)$$

For this reaction to take place a proton p must be converted into a neutron n through $p \rightarrow n + e^+ + \nu$, releasing a positron $e^+$ and a neutrino $\nu$. Such a conversion can only proceed through the weak interaction (see Subsection 3.1). Therefore, the rate of the reaction in Eq. (*2*) is very low, which makes the reaction the bottleneck of the pp-chain.

Once the deuteron d is formed, it very rapidly undergoes the reaction:

$$d + p \rightarrow {}^3\text{He} + \gamma. \tag{3}$$

There are two alternatives for the next step, leading to a branching of the pp-chain into the ppI- and ppII-chain. In the ppI-chain, occurring in 86% of the cases, two ${}^3$He nuclei fuse to a final ${}^4$He nucleus while two protons are released:

$$^3\text{He} + {}^3\text{He} \rightarrow {}^4\text{He} + 2\text{p} . \tag{4}$$

In the ppII-chain, occurring in 14% of the cases, a ${}^3$He nucleus fuses with a ${}^4$He nucleus creating a ${}^7$Be nucleus and thereby releasing a photon, $\gamma$:

$$^3\text{He} + {}^4\text{He} \rightarrow {}^7\text{Be} + \gamma . \tag{5}$$

In almost all cases this reaction is followed by the capture of an electron and emission of a neutrino $\nu$, thereby converting a ${}^7$Be nucleus into ${}^7$Li. This is followed by the capture of another proton, creating two ${}^4$He nuclei:

$$^7\text{Be} + e^- \;\rightarrow\; {}^7\text{Li} + \nu , \tag{6}$$

$$^7\text{Li} + p \;\;\;\rightarrow 2\ {}^4\text{He} .$$

Another branching into the ppIII-chain occurs in a very small percentage of cases with a total probability of only 0.02%. In this chain, reaction (*5*) is followed by the following sequence of reactions:

$$^7\text{Be} + p \;\;\rightarrow {}^8B + \gamma , \tag{7}$$

$$^8B \;\;\;\;\;\;\;\;\;\rightarrow {}^8\text{Be} + e^+ + \nu ,$$

$$^8\text{Be} \;\;\;\;\;\rightarrow 2\ {}^4\text{He} .$$

The net reaction of all three pp-chains

$$4\text{p} \rightarrow {}^4\text{He} + 2e^+ + 2\nu \tag{8}$$

leads to a transformation of four protons into a ${}^4$He nucleus releasing two positrons $e^+$, two neutrinos $\nu$ and a total energy of 26.73 MeV. A fraction of this energy is carried away by the neutrinos, which leave the star practically unhindered due to their negligible interaction with the solar material.

Fusion of hydrogen into helium may also be achieved through another sequence called CNO-cycle (Figure 11), which is notably different from the pp-chain:

$$^{12}C + p \rightarrow {}^{13}N + \gamma \,,$$

$$^{13}N \rightarrow {}^{13}C + e^+ + \nu \,,$$

$$^{13}C + p \rightarrow {}^{14}N + \gamma \,,$$

$$^{14}N + p \rightarrow {}^{15}O + \gamma \,,$$

$$^{15}O \rightarrow {}^{15}N + e^+ + \nu \,,$$

$$^{15}N + p \rightarrow {}^{12}C + \alpha \,. \tag{9}$$

In this sequence the C, N, and O nuclei only act as 'catalysts' and the net reaction of the CNO-cycle is again given by Eq. (*8*).

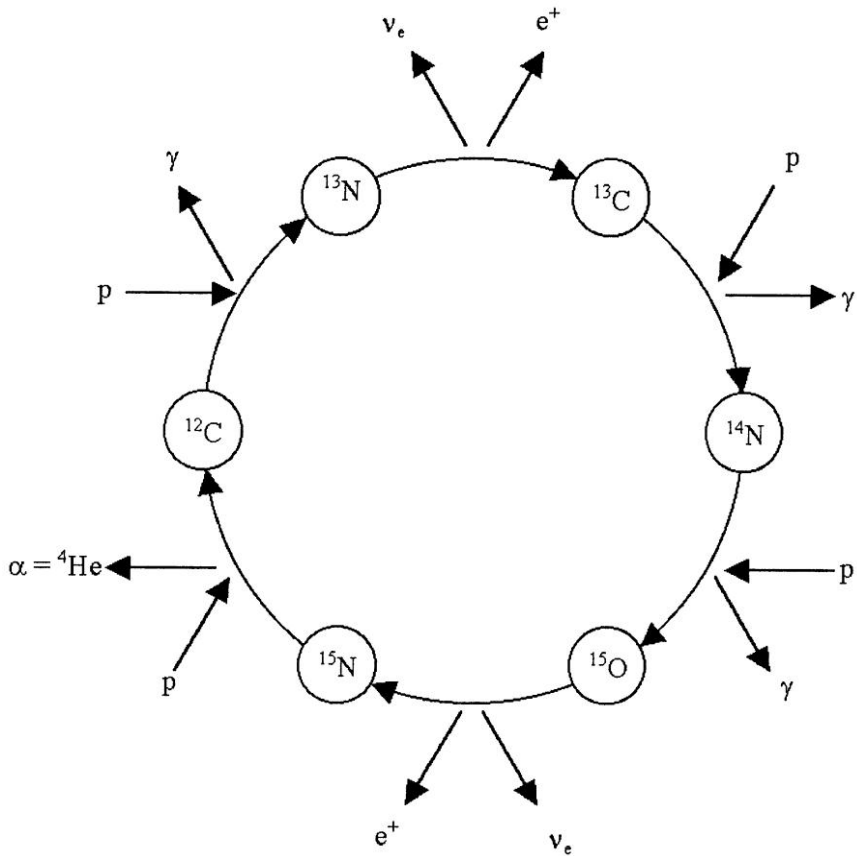

FIGURE 11. The CNO-cycle of hydrogen burning.

For main-sequence stars lighter than about two solar masses, the pp-chain dominates in hydrogen burning, whereas the CNO-cycle is favored over the pp-chain in stars that are more than twice as massive as the Sun.

In our Sun the pp-chain dominates over the CNO-cycle, producing about 98% of the total energy. A prerequisite for CNO-cycles is, of course, the existence of the elements C, N, O in the stellar plasma. Stars formed early in the Galaxy (first generation stars) contain only primordial elements and thus are not able to burn hydrogen through a CNO-cycle at all.

The Sun's temperature in the core is about $1.5 \times 10^7$ K, whereas the surface temperature is only about 5600 K. The Sun has already lived for 4.6 billion years, and will have enough hydrogen supply to live for about another 5 billion years. This leads to a long lifetime of about 10 billion years before its fuel is exhausted.

How can we obtain information from the Sun's core, where hydrogen burning takes place? The photons reaching us from the Sun are emitted from the solar surface. They have changed their energy enormously by many scattering processes on their way from the hot solar core to the relatively cool surface of the Sun. Therefore, in the visible region only the solar surface and not the solar interior can be observed. One possibility to obtain information about the Sun's core is helioseismology, i.e., by observing the vibration modes of the Sun. It was confirmed through helioseismology that our standard solar model is correct (Fiorentini and Ricci 2000, Bahcall *et al.* 2001a, Christensen-Dalsgaard 2001).

Another possibility is to observe the neutrinos that are set free in nuclear reactions of the pp-chain and CNO-cycle and reach the surface practically unhindered. Presently four neutrino detectors measuring solar neutrinos exist: HOMESTAKE (U.S.A), GALLEX (Italy), (SUPER)KAMIOKANDE (Japan) and SNO (Canada). All the above neutrino detectors are underground in order to shield out the cosmic rays that would give unwanted background signals in the neutrino detectors. The existing solar neutrino detectors measure only about 1/3 to 1/2 of the electron neutrino flux compared to the value calculated from the standard solar model (Bahcall 2000). This discrepancy is called the solar neutrino problem (Bahcall 1989, 1999). Possible problems both with the neutrino measurements and with our standard solar model have been ruled out. Recently, at SNO it was possible to observe not only the solar electron neutrinos, but also the $\mu$- and $\tau$-neutrinos in the same experiment (Heger 2001, SNO collaboration 2002a, SNO collaboration 2002b). The measured total neutrino flux agrees with the value expected from the standard solar model. The solution to the solar neutrino problem implies some new physics by the introduction of the so-called neutrino oscillations (see Subsection 8.2 in Chapter 8, Volume 1). Through such oscillations, the electron neutrinos $\nu$ emitted in the solar core by the nuclear reactions given in Eqs. (*2*), (*6*) and (7) can change into other types of neutrinos. Thus, mainly $\mu$-neutrinos emerge on their way from the core of the Sun to the detector. Experiments measuring electron neutrinos thus show a smaller flux than initially emitted in the solar core (Bahcall 2001, Bahcall *et al.* 2001b, Fiorentini *et al.* 2001). This physical picture is the culmination of about 40 years of solar neutrino detection and research.

Recently, these findings are combined with the results of the so-called atmospheric neutrino anomaly, where $\mu$-neutrinos generated in pion decays oscillate over into, mainly, $\tau$-neutrinos. Also terrestrial experiments performed with neutrino fluxes produced either at nuclear power plants or with accelerators provide substantial information on the mixing pattern among the different neutrino species. Searches for possible oscillations into further light neutrino flavors, which would be of clear cosmological significance, did not provide a clear answer to the question of their existence, yet. (See Subsection 8.2 in Chapter 8, Volume 1)

## 4.3   Helium burning: nucleosynthesis of carbon and oxygen

The fusion of protons into helium continues until the star has exhausted its hydrogen. When this happens, the star undergoes a gravitational collapse and the temperature rises to about a few times $10^8$ K in the core of the star, which makes the fusion of helium into heavier nuclei possible. In the first reaction of helium burning the fusion of two $^4$He nuclei creates the $^8$Be nucleus. However, the $^8$Be nucleus has an extremely short mean life of only $10^{-16}$ s, before it decays back again to two $^4$He nuclei. This process is in equilibrium, where the rate of production equals the rate of destruction of the $^8$Be nucleus:

$$^4\mathrm{He} + {}^4\mathrm{He} \leftrightarrow {}^8\mathrm{Be} \,. \qquad (10)$$

The $^8$Be just produced can, however, capture another $^4$He nucleus creating the $^{12}$C nucleus through the reaction:

$$^8\mathrm{Be} + {}^4\mathrm{He} \rightarrow {}^{12}C + \gamma \,. \qquad (11)$$

The reactions (*10*) and (*11*) are called the triple-alpha reaction, because three $^4$He nuclei or alpha particles are necessary for the creation of $^{12}$C. This reaction can only create carbon in appreciable amounts because of the existence of a resonance in $^{12}$C at the relevant energy for helium burning. Through this resonance the reaction (*11*) is enhanced by many orders of magnitude.

In helium burning about half of the carbon nuclei produced are converted to oxygen nuclei $^{16}$O by the capture of another $^4$He nucleus:

$$^{12}C + {}^4\mathrm{He} \rightarrow {}^{16}O + \gamma \,. \qquad (12)$$

Further captures of helium nuclei $^4$He by oxygen nuclei $^{16}$O occur only to a much lesser extent and therefore helium burning comes to an end after the creation of $^{12}$C and $^{16}$O.

Carbon and oxygen are the two most important elements for carbon-based life. Carbon is needed for the complex nuclei of the DNA and proteins, whereas oxygen is needed even for water. Interestingly enough, these two elements are extremely fine-tuned with respect to the nuclear force. If the strength of this force were 0.5% larger or smaller, the average abundance of carbon or oxygen in our Universe would be reduced by more than two orders of magnitude. This would make the existence of carbon-based life in our Universe very improbable (Oberhummer *et al.* 2000; Schlattl *et al.* 2004).

Outside the stellar core burning helium, hydrogen burning continues in a shell around the core. If the initial mass of a star is less than about 8 solar masses no more burning phases will take place after helium burning and nuclear burning stops. A white dwarf with a surrounding expanding planetary nebula will be the endpoint of the star's life. Charbonnel *et al.* (1999) and Marigo (2000) review the chemical yields in light and intermediate-mass stars between 0.8 and 8 solar masses.

Two interesting effects happen in helium burning of low mass stars. Firstly, stars with less than about 2 solar masses undergo a **core He flash** instead of igniting stable He burning after the H burning phase. This is because the cores of lower mass stars are more dense than those of higher mass stars. For stars with less than about 2 solar masses, the contracted He core is so dense that it cannot be described as an ideal gas. Rather, it is a degenerate gas, in which pressure is only depending on density but not on temperature. Once the triple-$\alpha$ reaction, being very efficient at high density, is ignited, a thermonuclear runaway ensues because the rising temperature does not rise the pressure and therefore does not cause an expansion of the burning zone. Thus, the usual selfregulation mechanism of hydrostatic burning is not working anymore and He burning proceeds quickly to high temperature. Very high temperatures lift the degeneracy of the gas and its equation-of-state becomes temperature dependent again very suddenly. This causes an explosive expansion of the outer core, also ejecting the outer layers of the star as a planetary nebula.

The second phenomenon occurs in stars between 2 and 8 solar masses, the so-called Asymptotic Giant Branch (AGB) stars. Regular He burning takes place in the core of these stars in their red giant phase. With the exhaustion of He in the center of the star, the burning zone moves outward and becomes a burning shell. Thus, there are two shells burning, a H-burning shell and a He-burning shell. The He-burning shell is very thin and does not generate sufficient energy to balance the mass layers on top of it through radiation pressure. This squashes the shell more and more. Because of the non-linear dependence on density of the triple-$\alpha$ rate, which is actually two reactions one after another, the energy release will considerably increase but still not be enough to expand and selfregulate the shell against the pressure from the surrounding layers. Further contraction enhances the triple-$\alpha$ rate nonlinearly and so on. Although the gas is not degenerate, a similar thermonuclear runaway as in the degenerate case occurs. When a critical temperature is reached, enough energy is released to explosively expand the shell against its surroundings. This rapid expansion, the **He-shell flash**, is so strong that it also blows out the H-burning shell. Due to the expansion, the density drops and the triple-$\alpha$ reaction ceases. Quickly, the star contracts again, the outer material settles, and first the H-burning and then the He-burning shell is ignited again. This sets the stage for another such cycle. AGB stars undergo a large number of such pulses, where the thermonuclear runaway phase with the flash lasts only a few hundred years whereas the time between pulses is a hundred to a thousand times longer. Oscillations and vibrations are induced into the stellar plasma by these pulses, leading to increased mass loss from the surface of the star. AGB stars have strong stellar winds which considerably decreases their total mass during their evolution. The shell flashes have another important impact: they cause large convection zones, mixing the plasma constituents across large distances within the star. This is important for the production of the s-process nuclei (see Section 4.5.1).

He-shell flashes only occur in low-mass stars because they are caused by thin He-burning shells and the size of the shells scale with the stellar mass.

## 4.4 Advanced burning stages

In a massive star with more than 8 solar masses, the next stage after helium burning is carbon burning. This starts when the carbon/oxygen core has shrunk so that the temperature at its center has reached about $5 \times 10^8$ K. Then two carbon nuclei fuse together creating $^{20}$Ne or $^{23}$Na nuclei:

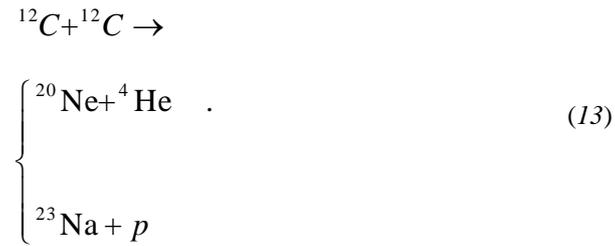

$$^{12}C + ^{12}C \rightarrow$$

$$\begin{cases} ^{20}\text{Ne} + ^4\text{He} \\ \\ ^{23}\text{Na} + p \end{cases} \qquad . \qquad (13)$$

The next stage is neon burning starting at $10^9$ K, in which photons first disintegrate $^{20}$Ne and liberate $^4$He, which in turn reacts with the undissociated $^{20}$Ne to build up $^{24}$Mg and further nuclei:

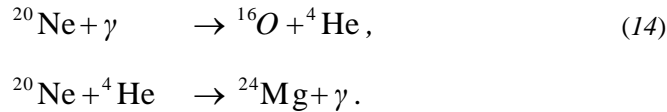

$$^{20}\text{Ne} + \gamma \quad \rightarrow \, ^{16}O + ^4\text{He} \, , \qquad (14)$$

$$^{20}\text{Ne} + ^4\text{He} \quad \rightarrow \, ^{24}\text{Mg} + \gamma \, .$$

Oxygen burning occurs when the temperature reaches $2 \times 10^9$ K, the most important reaction being the one producing $^{28}$Si:

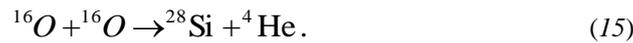

$$^{16}O + ^{16}O \rightarrow \, ^{28}\text{Si} + ^4\text{He} \, . \qquad (15)$$

The final stage is reached at a temperature of $5 \times 10^9$ K, when silicon burning begins. At this high temperature a series of reactions takes place beginning with the photodisintegration of $^{28}$Si:

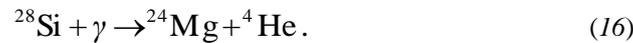

$$^{28}\text{Si} + \gamma \rightarrow \, ^{24}\text{Mg} + ^4\text{He} \, . \qquad (16)$$

Then the released $^4$He nuclei build up heavier nuclei by successive capture reactions:

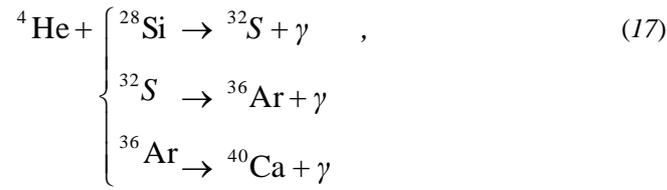

$$^4\text{He} + \begin{cases} ^{28}\text{Si} \rightarrow ^{32}S + \gamma \\ ^{32}S \rightarrow ^{36}\text{Ar} + \gamma \\ ^{36}\text{Ar} \rightarrow ^{40}\text{Ca} + \gamma \end{cases} \quad , \qquad (17)$$

and so on. At such high temperatures capture and photodisintegration reactions are in equilibrium. In this so-called nuclear statistical equilibrium (NSE) the knowledge of individual reactions or reaction rates is not important anymore to calculate the abundances. The produced abundances only depend on the temperature and density of the plasma and the nuclear binding energies (Iliadis 2007, Boyd 2008). The result of this series of photodisintegration and capture reactions is the steady build-up of heavier elements up to the elements grouped around iron (Hix and Thielemann 1998), with $^{56}$Ni preferentially produced because it is the nucleus with the highest binding energy and having equal number of protons and neutrons.

The sequence of stellar burning is terminated when the core of the star is largely composed of elements in the mass region of nickel and iron, because no more energy is to be gained from further nuclear reactions. As soon as the energy produced is not enough to maintain the hydrostatic equilibrium, the core cannot support the outer layers anymore and it begins to collapse due to its gravitation, leading to a core-collapse supernova (see Subsection 4.1.3).

In a core-collapse supernova explosive nucleosynthesis also takes place through the outward proceeding shock wave (Thielemann *et al.* 2001a), modifying the elemental abundance pattern of the outer layers of the pre-supernova star. This **explosive burning** of the C-, Ne-, O- and Si-layers in the star mainly leads to modifications of the abundances in the region from Ca to Fe (Rauscher *et al.* 2002). Photodisintegration of heavy nuclei also leads to the production of proton-rich stable nuclides, the so-called p-nuclides (see also Subsection 4.5.2). The strong neutrino emission caused by the formation of a neutron star in the core collapse influences nucleosynthesis in the deepest, barely ejected layers of the star as well as in the outer layers (Sections 4.5.2 and 4.5.3).

Nucleosynthesis of massive stars is reviewed by Rauscher *et al.* (2002), Woosley and Heger (2007). An overview of stellar nucleosynthesis including hydrogen, helium, neon, silicon, and explosive burning as well as the basics of the s- and r-process are given by Rauscher and Thielemann (2001). Explosive burning and the s- and r-processes are also introduced below.

## 4.5    Nucleosynthesis beyond Fe

As we have seen in the preceding sections, stellar burning phases only lead to the production of nuclei up to Fe. A review by National Research Council of the National Academies identified 11 key questions to be addressed in science in the next decade (Turner *et al.* 2003). Ranked 3 on the list is "How were the elements from Fe to U made?" Although the ground to answer this question has been laid by Burbidge *et al.* (1957), Cameron (1957) and much progress has been made since then, there remain a number of problems regarding the astrophysical sites of certain nucleosynthesis processes and also concerning the properties of certain, highly unstable nuclei in such processes. In the following sections we give a brief summary of the current knowledge of how elements beyond Fe were synthesized.

### 4.5.1   The main and weak s-process components

Due to the lack of a Coulomb barrier, the most likely process for the formation of elements heavier than those grouped around iron is neutron capture. If a supply of neutrons is available, they can accrete by sequential neutron captures on a 'seed nucleus' in the region of iron to build up neutron-richer nuclei. As the neutron number of the nucleus increases, it will become unstable to $\beta^-$ decay, transforming a neutron into a proton in the nucleus and emitting an electron and an antineutrino. Successive neutron captures, interspersed by $\beta^-$ decays build up many, but not all of the heavier stable nuclei.

There are two basic time scales in this scenario of heavy-element nucleosynthesis by neutron captures: (1) the beta-decay lifetimes, and (2) the time intervals between successive captures that are inversely proportional to the neutron capture reaction rates and the neutron flux. If the rate of neutron capture is slow compared to the relevant $\beta$ decays, the synthesis path will follow the bottom of the stability valley very closely. On the other hand, if the rate of neutron capture is faster than the relevant $\beta^-$ decays, highly neutron-rich nuclei will be formed. After the neutron flux has ceased, those nuclei will be transformed to stable nuclei by a series of $\beta^-$ decays. The above two processes are called s- and r-process, respectively, according to their slow or rapid rate of neutron capture. The observed abundances of nuclei in the solar system, especially in the regions of closed-shell nuclei, suggest that the s- and r-processes contributed more or less equally to the formation of the elements above the iron region (see Figure 12).

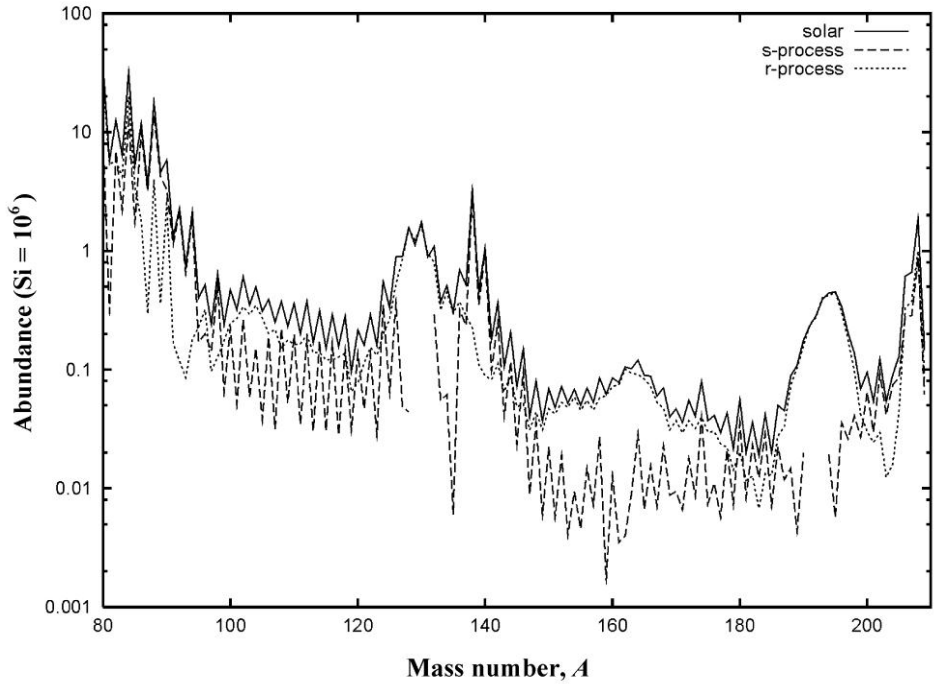

FIGURE 12. Contribution of s- and r-process to the solar abundances of the isobars for heavy elements (p-isotopes cannot be seen in this figure because of their very small abundances). Solar system abundances are measured (Anders and Grevesse 1989), s- and r-abundances are calculated. The peaks in the solar abundances around mass numbers $A = 88, 138, 208$ are formed in the s-process, whereas the broader companion peaks shifted to slightly lower mass number are r-process peaks.

Two important reactions provide neutrons for the s-process:

$$^{13}C + {}^4\mathrm{He} \quad \rightarrow {}^{16}O + n\,, \qquad (18)$$

$$^{22}\mathrm{Ne} + {}^4\mathrm{He} \quad \rightarrow {}^{25}\mathrm{Mg} + n\,.$$

The reaction on $^{13}$C is much more efficient in releasing neutrons because it is strongly exothermic, contrary to the second reaction. However, $^{13}$C normally does not occur in He-burning zones whereas $^{22}$Ne does. This proved to be a longstanding problem in complete stellar simulations of s-process nucleosynthesis.

Early observations (see Burbidge *et al.*, 1957) had already found Tc on the surface of AGB stars. Since Tc isotopes are short-lived compared to the age of such a star, they had to be produced in that star and brought up to the surface. Only in recent years, sophisticated stellar models were able to follow the complicated convection and nucleosynthesis processes inside AGB stars with sufficient accuracy to confirm them as the production sites. He-shell flashes (see Section 4.3) and the mixing brought about by them turned out to be the key (Busso *et al* 2001, Boothroyd 2006). Such a flash can mix down protons from the unburnt outer layer of the star which can then be used to produce $^{13}$C by proton capture on $^{12}$C and subsequent $\beta$-decay of the resulting $^{13}$N. With this $^{13}$C neutrons can be very sufficiently released even during the interpulse phase. Additionally, the reaction on $^{22}$Ne can release further neutrons during the high temperature phase of the flash itself. Thus, the nuclides in the stellar plasma are irradiated with neutrons in bursts over millennia. The large convection zones appearing in the flash phase bring up the newly synthesized material to the surface.

In the manner described above, AGB stars produce the majority of the s-process nuclei, the so-called main component. It was known for a long time that there must be a second site of s-processing, producing light s-nuclei. Because they exhibit smaller abundances than those of the main component, this was called the weak component. Such a weak s-process is found in massive stars (i.e. stars with more than 8 solar masses), where capture of $^{4}$He by $^{22}$Ne is the main neutron source. Massive stars reach higher temperatures than AGB stars already in their late evolution stages which releases neutrons. Even more can be released during explosive burning, when the temperature rises due to the supernova shock wave passing through the outer layers of the star. Because of the inefficiency of the $^{22}$Ne neutron release and the short timescale, one cannot proceed much beyond Fe by this mechanism in massive stars.

For nucleosynthesis of the heavy elements through the s-process in both AGB and massive stars, there already must be nuclei present in the iron region, which were produced in previous generations of stars. Thus, the s-process will be stronger in stars formed more recently than in older stars containing less heavy elements.

### 4.5.2 Explosive nucleosynthesis in the outer layers of a massive star

Since the build-up of nuclei in the s- (and r-) process follows the neutron-rich side of the stability valley (editors' note: see, e.g., Figure 13 in Chapter 5, Volume 1), 32 proton-rich isotopes cannot be produced in either process. These so-called **p-nuclides** occur naturally but with abundances many orders of magnitude lower than the other nuclides. The hypothetical process synthesizing these nuclides was termed **p-process** and several models have been suggested. The commonly favored one is photodisintegration of pre-existing nuclei in the Ne/O shells of massive stars. When a supernova shockfront is passing through these layers, the high temperatures of 2-4 GK enable photodisintegrations, starting by $\gamma$-induced emission of several neutrons, leading to proton-rich nuclei. The photodisintegration path can branch when proton or $\alpha$ emission becomes more favorable than neutron emission in such proton-rich nuclei. The bulk of p-nuclides can be explained in such a model but some problems remain (Rauscher *et al.* 2002, Arnould and Goriely 2003, Boyd 2008). Especially the production of the light p-isotopes, in particular $^{92,94}$Mo and $^{96,98}$Ru, is not understood. Among the p-nuclei they have the by far highest abundance but cannot be made concurrently with the others. It remains an open question whether the stellar models have to be revised or an additional production mechanism has to be invoked for these light p-nuclei.

The neutrino flux of a core-collapse supernova is high enough to contribute to the nucleosynthesis of certain rare elements and isotopes, even in the outer layers of the star. In this so-called ν-process, inelastic scattering of a neutrino leads to formation of an excited daughter nuclide, which then decays by particle emission. This process can contribute significantly to the production of light ($^{11}$B, $^{19}$F) and heavy ($^{138}$La, $^{180}$Ta) isotopes (Woosley and Weaver 1995, Heger *et al.* 2005).

### 4.5.3 *Explosive burning in the deep layers of a massive star*

In addition to s-process nucleosynthesis, about half of the nuclides beyond Fe are produced through rapid neutron captures on short timescales in the r-process. The site of the r-process is controversial. Mostly favored are core-collapse supernovae where appropriate r-process conditions are thought to be found close to the region of neutron star formation. These innermost layers, which are barely ejected, move outwards within a strong neutrino flux, driving the material to become very neutron rich. With a high neutron density, neutron captures can proceed much faster than β-decays and produce very neutron-rich nuclei far from stability. Through simultaneously occurring captures, photodisintegrations with neutron emission, and β-decays heavier elements are synthesized within a few seconds. When the ejected material cools down, those highly unstable nuclei decay back to stability, thus supplying the needed fraction of heavy elements. While the s-process is confined to the region up to Bi, the r-process is thought to also reach the region of fissionable nuclei and produce natural, longlived elements such as U. The endpoint of the r-process path is highly debated since it depends on fission barriers of very neutron-rich, heavy nuclei, for which there is no consensus among theoretical models, yet.

The conditions in those innermost regions of a core-collapse supernova are closely linked to the working of the explosion mechanism. Since the latter is not yet fully understood, it is not yet clear whether the required conditions can actually be established. Therefore, a number of alternative scenarios is still discussed, such as jet outflows from asymmetrically exploding stars. The search for the site of the r-process remains a major focus of research.

Recently, an additional nucleosynthesis process in the deep layers of the exploding star has been suggested (Fröhlich et al. 2006). It was discovered that the combined flux of neutrinos and antineutrinos from the emerging, hot neutron star initially creates very proton-rich conditions before the matter becomes neutron-rich at later times and/or larger radii. The high temperature and density environment gives rise to rapid proton captures, thus synthesizing nuclei beyond Fe but on the proton-rich side of stability. A small number of neutrons is required to speed up the matter flow to heavier elements and these are produced by antineutrino captures on protons. . The νp-process could perhaps explain the surprisingly high abundance of Sr, Y, Zr found in very old stars (Travaglio *et al*, 2004; Frebel *et al*, 2005). Again, the details of this so-called νp-process and how efficient it can produce elements beyond Fe depends strongly on the conditions in the deep layers of the exploding star and the explosion mechanism. Among the suggested alternative scenarios are wind outflows from the accretion disks around black holes formed by core collapse of very massive stars (Surman *et al.* 2006). These are also thought to be the cause of so-called γ-ray bursts which are the most energetic phenomena observed in our Universe today (MacFadyan and Woosley 1999; Mészáros 2006).

## 4.6   Nucleosynthesis by spallation

The light and fragile elements lithium, beryllium, and boron (LiBeB) are not primarily produced in primordial or stellar nucleosynthesis. In fact, the abundance curve in Figure 13 shows a huge dip (almost a gap, actually) for the mass numbers 8-11, reflecting the scarcity of LiBeB-nuclei in the solar system. Only the nuclide $^7$Li can be produced both in primordial (see Subsection 3) and in stellar nucleosynthesis (see Subsection 4.2), whereas the isotopes $^6$Li, $^9$Be, $^{10}$B and $^{11}$B are almost pure spallation products of heavier elements.

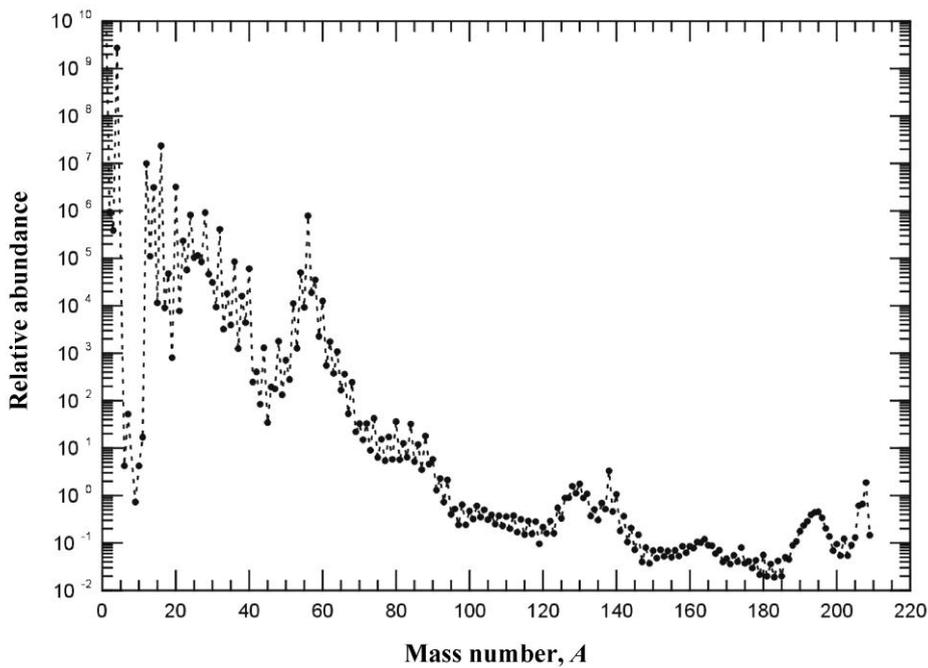

FIGURE 13. Relative solar abundances by mass number. Data are from Anders and Grevesse 1989. The abundances are arbitrarily renormalized to yield a value of $10^6$ for silicon.

The high-energetic Galactic Cosmic Rays (GCRs) originate probably from supernovae (Erlykin and Wolfendale 2001). GCRs consist mainly of fast-moving bare hydrogen and helium nuclei and, to a lesser amount, of carbon, nitrogen and oxygen nuclei (CNO) nuclei. Hydrogen and helium nuclei of interstellar clouds can spall the CNO-nuclei in flight of the fast GCRs. Therefore, the GCRs are by about a million times enriched in LiBeB-nuclei compared to the solar system abundance.

The most plausible origin of the main bulk of LiBeB-nuclei is that hydrogen and helium nuclei of GCRs hit and spall CNO-nuclei contained in interstellar clouds. However, this process alone seems unable to produce LiBeB at the observed level. Therefore another production site of LiBeB-nuclei has been proposed. This invokes in-flight fragmentation of carbon and oxygen nuclei by collision with hydrogen and helium nuclei in interstellar clouds. The sites of this process are mainly the surroundings of massive stars, which are able to furnish freshly synthesized carbon and oxygen nuclei and accelerate them via shock waves. Finally, spallation through neutrinos in supernova explosions also produces the nuclides $^7$Li and $^{11}$B (Hartmann *et al.* 1999).

A review of nucleosynthesis by spallation is given by Vangioni-Flam *et al.* (2000).

# 5    NUCLEOSYNTHESIS IN BINARY STAR SYSTEMS

## 5.1    General considerations

Observations reveal that more than half of the known stars are associated in binary or multiple systems. In many such systems, the stars are so well separated that there is negligible influence on each other's evolution. However, when the constituents of such a system are close, a range of interesting physical phenomena can ensue. Mass can be transferred from one star to the other when the donating star becomes so large that its atmosphere extends beyond the limiting region where the gravitational attraction of the two stars is equal. Material is then escaping from the gravitational well of the donor and flowing into the attractive field of the other object. It can be shown that there is a single point of gravitational equilibrium between the two objects, through which the mass transfer will occur. The volume around a star enclosed by the equipotential surface containing this point is called the Roche lobe. A star can fill or exceed its Roche lobe when it either increases its radius in the normal course of evolution (e.g. by becoming a Red Giant), or when the orbital separation of the two objects is decreasing due to the emission of gravitational waves. The latter situation corresponds to shrinking their respective Roche lobes. The inflow onto the surface of the companion can alter its surface properties, which is manifested, for instance, in the observed spectra. The further evolution might even lead to powerful explosions. A broad review of such phenomena is given by Warner (1995).

From the point of view of nucleosynthesis, explosive burning in such binary systems proves to be especially important as nuclides created in such an environment can be injected into the interstellar medium by the explosion. Furthermore, the nucleosynthetic signature will be different from stellar hydrostatic and explosive scenarios due to the different initial and burning conditions. Thus, different isotopic ratios can be obtained and, in some cases, it is possible to synthesize nuclides not accessible in other sites. In the following, we therefore focus on outlining only such scenarios which might have an impact on the abundances of elements observed in our Galaxy and the Universe. It has to be emphasized, however, that it is not yet possible to model all details of the explosive processes, due to the complexity of a multi-dimensional hydrodynamic problem including scales differing by many orders of magnitude. The problems are similar to those occurring in core-collapse supernova models. In scenarios where neutron stars are involved, additional uncertainties in the nuclear equation of state enter. As a result, a detailed, quantitative understanding of some of the processes is still lacking. Nevertheless, the basic features of all the processes can be described qualitatively in simplified models.

## 5.2 Novae

Historically, the Nova phenomenon is one of the oldest observed. Already astronomers of ancient civilizations were aware of sudden appearances of bright stars in the sky, at a position where no star had been observed before. This was termed 'nova' as in 'new star'. Modern astronomy was able to identify nova progenitors. These are faint stars (mostly undetectable by the naked eye) which suddenly increase their brightness by several stellar magnitudes (i.e. their energy output increases by a factor of 100 - 100 000) within a few days. The star returns to its initial brightness after several months. In rare cases the same star experiences another nova outburst within several years. This is called a recurring nova. In general, classical novae reappear only every 100 to 10 000 years.

The modern explanation of the Nova phenomenon invokes a binary system with a regular star in its early burning phases and a white dwarf (Fujimoto 1982, Truran 1984, Gehrz *et al.* 1998, Starrfield 1999). The latter is too faint to be directly observed. Such a system can have originated from two low-mass stars, one of them slightly more massive than the other. More massive stars evolve faster and thus the heavier star already underwent its complete stellar evolution and left a white dwarf as a remnant, while the other star has not yet even completed its early phases. From the atmosphere of the main sequence or red giant star hydrogen-rich material is flowing across the Roche surface. It is captured by the gravitation of the white dwarf and forms an accretion disk around it. Subsequently, the gas is accreted on the surface of the companion, forming a thin layer. Close to the surface the gas becomes degenerate, delaying the onset of hydrogen burning. As soon as the necessary density and temperature is reached, hydrogen burning ignites explosively, pushing outwards and ejecting part of the outer accreted layer. With the sudden expansion, temperature and density are dropping again and the burning ceases. It can reoccur as soon as a sufficient amount of new material has been accreted which usually takes several thousand years at typical accretion rates of less than 1 Earth mass per 10 years.

In terms of nucleosynthesis it should be obvious that nova ejecta are mainly hydrogen and that they more or less retain the surface composition of the accompanying star. However, it was found observationally that they are also strongly enriched in C, N, and O. This has been a puzzle for over two decades (Rosner *et al.* 2001). In the thermonuclear runaway triggered by hydrogen ignition the main reaction sequences are those of the CNO-cycles. However, they can account neither for the observed enrichment nor for the energy production required for a fast nova burst. Currently, it is thought that material from the surface of the white dwarf might be dredged up and efficiently mixed into the burning layer and the outer zones. This would increase the CNO energy production and also provide a mechanism for the enrichment. However, to date one-dimensional and multi-dimensional simulations have not been able to account for the observations (Starrfield *et al.* 2000, Kercek *et al.* 1999). Nevertheless, it seems that novae are 'digging up the ashes' of previous stars and distributing them throughout the Galaxy. They are considered to be the major sources of $^{15}$N, $^{17}$O, and $^{13}$C, and to have minor contributions to a number of additional species, mainly $^{7}$Li and $^{26}$Al. For current reviews on the nucleosynthetic contribution of novae, see Jordi and Hernanz (2007), Jordi and Hernanz (2008)

## 5.3   Type Ia supernovae

Given a binary system consisting of a white dwarf and a companion star as in the nova case, the accretion rate on the surface of the white dwarf is essential for the further development. Low accretion rates lead to a nova as described above. When the accretion rate exceeds about $10^{-8}$ solar masses per year, hydrogen can be quiescently burned during accretion and the burning products will sediment on the surface of the white dwarf, forming a He layer. The details on the further fate of the object are complicated and have not been fully simulated yet. Basically two ways of explosion can be envisioned. In the first, the ignition of the He layer leads to a thermonuclear runaway, this time mainly burning via the triple-$\alpha$ reaction, not depending on CNO elements, contrary to novae. Nevertheless, the resulting explosion would lead to an expulsion of the outer layers, like in a nova, only much more powerful. The second type of explosion occurs when the accretion rate is even higher, about $10^{-6}$ solar masses per year. The energy released by the accretion and by burning layers heats the C/O core of the white dwarf sufficiently to ignite core C burning. This is followed by a complete disruption of the white dwarf because the nuclear energy exceeds the gravitational binding energy (Leibundgut 2001a,b).

The latter is the currently most widely accepted scenario explaining SN Ia (Nomoto *et al.* 1984, Hillebrandt and Niemeyer 2000). It is encouraging that observations can confirm also the basic nucleosynthesis features expected. In each explosive event a quantity of material of the order of 0.6 - 0.8 solar masses is produced in the $^{56}$Ni region, which later decays to its stable isobars. Thus, the larger part of the Fe found in the Solar System stems from SN Ia. In addition, some intermediate elements like Mg, Si, S, Ca are also produced.

The details of the shockwave propagation and explosive burning are not fully understood yet. However, the total energy production remains robust due to the fact that the initial white dwarf mass is always close to the Chandrasekhar limit of 1.4 solar masses. This fact, in combination with the observational evidence that there are no compact central objects (i.e. neutron stars) found in SN Ia remnants, puts a strong constraint on the achieved explosion energy. This is why SN Ia are thought to be excellent standard candles with very low variation in the effective energy output (Dominguez *et al.* 2001, Leibundgut 2001a,b). SN Ia as standard candles are important tools to measure astronomical distances although there is no firm theoretical grounding of the Philips relation (see subsection 2.1.4), yet.

## 5.4  X-ray bursts and the rp-process

Bursts have been observed not only in the optically visible frequency range of light. Powerful, brief bursts of X-rays are also observed throughout the Galaxy. A binary system is suggested to be responsible for one subclass of these bursts. Their duration is from several seconds up to minutes and they show a fast rise and a slowly decaying tail. In the assumed scenario, a main-sequence star and a neutron star are orbiting the common center-of-mass. As in the nova and SN Ia case, material is flowing from the atmosphere of the companion star and is accreted on the surface of the compact object. Due to the increased gravitational field of the neutron star in comparison with a white dwarf, the thermonuclear runaway after ignition of hydrogen burning can proceed differently from the one in novae and supernovae (Taam *et al.* 1996, Schatz *et al.* 1998, Wiescher and Schatz 2000, Boyd 2008). Hydrogen and subsequently also helium, burn explosively at higher temperatures as before. First, the so-called hot CNO-cycle (also found in certain massive stars) is established, followed by further CNO-type cycles beyond Ne. The energy production from these cycles leads to a break-out to further CNO-type cycles beyond Ne at temperatures surpassing $4 \times 10^8$ K. The additional cycles generate additional energy, further increasing the temperature. In the next stage of the ignition process He is also burned in the triple-$\alpha$ reaction. The CNO-type cycles break up towards more proton-rich nuclides by ($\alpha$,p) and (p,$\gamma$) reactions.

Finally, the rp-process (rapid proton capture) sets in. In the rp-process, similar to the r-process for neutrons, proton captures and photodisintegrations are in equilibrium. Thus, the abundances within an isotonic chain are only determined by temperature, density and the proton separation energy of a nucleus. The time scales of flows from one chain into the next are given by beta-decay half-lives. The reaction path basically follows the proton dripline. The processing of matter is hampered at nuclides with long half-lives, the so-called waiting points, which determine the processing time-scales. A final endpoint of the rp-process path was found in a closed reaction cycle in the Sn-Sb-Te region, due to increasing instability against $\alpha$ decay of heavy proton-rich nuclides (Schatz *et al.* 2001).

Time-dependent calculations showed that the structure of type I X-ray bursts can be explained by the energy generation of the proposed processes. Regarding nucleosynthesis, it is being discussed whether a fraction of or all light p-nuclei (see Section 4.5.2) originate from X-ray bursters. Since the rp-process synthesizes very proton-rich nuclei, they would decay to p-nuclei after the burning ceased. Since it proved to be problematic to synthesize the light p-nuclei with mass numbers $A<110$ in the $\gamma$-process by photodisintegration (Section 4.5.2), the rp-process provides a compelling alternative to such models, especially after the discovery of its endpoint, preventing the production of nuclides with mass numbers $A > 110$. Although it was shown that the isotopes in question can be produced in large quantities, it is still speculative whether any of the material is ejected. There could be a very small fraction of material lost from the outer atmosphere of the accreted layer on the neutron star surface but the rp-process burning takes place further down. Therefore, some convection has to be invoked to bring the freshly produced nuclei to the outer layers. Comparatively small amounts of ejecta would be sufficient to explain solar p-abundances but detailed hydrodynamic studies of the burning, convection, and possible ejection are still needed.

## 5.5 Neutron star mergers

Another interesting binary system is that of two neutron stars. Such systems are known to exist; four have been detected by now. They can be created when two massive stars complete their stellar evolution and both explode in a core-collapse supernova, each leaving a neutron star behind. In such a configuration, the system loses angular momentum by emission of gravitational radiation (Taylor 1994) and the two neutron stars spiral inwards. At time-scales of $10^8$ y or less, the objects collide at their center-of-mass. Such a merger can lead to the ejection of neutron-rich material (Rosswog *et al.* 1999, Ruffert and Janka 2001). Since this material is even more neutron-rich than the deep, high-entropy layers thought to be a possible site of the r-process in core-collapse supernovae, nucleosynthesis in decompressed neutron star matter could be a viable alternative site for the r-process (Freiburghaus *et al.* 1999). Detailed hydrodynamic calculations coupled to a complete r-process reaction network have not been undertaken yet, but parameterized r-process studies indicate the possibility that such mergers could even account for all heavy r-process matter in the Galaxy. However, detailed Galactic Chemical Evolution models (Argast *et al.* 2004) show that neutron star mergers, occurring at late time in the life of a galaxy, cannot account for the r-process nuclei found in very old stars (Sneden *et al.* 2000, Cayrel *et al.* 2001, Frebel *et al.* 2005). Therefore, there may be several sites producing r-process nuclei, perhaps similar to the two components of the s-process (but occurring in different sites than the s-process, of course).

# 6  THE ABUNDANCE OF ELEMENTS IN THE UNIVERSE

## 6.1  Experiments and observations

The essence of every science is the ability to validate theories by comparison to experiments. The ability to perform guided experiments is somewhat limited in astrophysics because of the large scales and extreme conditions involved. Nevertheless there is a wealth of data to be exploited. Atomic transitions and certain nuclear reactions (Käppeler *et al.* 1998) can be studied in the laboratory with current methods.

Since the s-process (Section 4.5.1) involves mainly stable nuclei, an important experimental contribution is the one studying neutron capture at low energy. Special focus is put on reactions at **s-process branchings**. For the majority of unstable nuclei reached through the s-process, β-decays are much faster than the neutron captures. At several places in the nuclear chart, however, the s-process path encounters long-lived nuclei for which the decay rate becomes comparable to the neutron capture rate. This leads to a splitting of the path: a fraction of the s-process flow proceeds through neutron capture, the other through the decay, bypassing certain isotopes. A comparison of abundances of nuclei reached through one or the other branch provides information on the relation between decay and capture rate. Due to the different temperature dependence of capture and decay, the branching ratio is dependent on temperature, becoming a sensitive s-process "thermometer". The capture rate also depends on the neutron density and thus such a branching can also be used as neutron pycnometer. This yields detailed information on the conditions inside an AGB star. To be useful, the neutron capture cross sections have to be known with an accuracy of better than 1% below about 50 keV neutron energy. This has been achieved for some target nuclei but remains a challenge for others. High-resolution time-of-flight experiments are most promising to give the required accuracy and provide cross sections within the required energy range. The cross sections have to be converted to reaction rates to be applied in astrophysical models. The Karlsruhe group (Käppeler *et al.* 1989) has had a leading role in directly determining rates by using a neutron spectrum created by the $^7$Li(p,n)$^7$Be reaction. Through this trick, the resulting energy distribution of the released neutrons is very similar to the one of neutrons in a stellar plasma with thermal energy of 25 keV, coinciding with s-process conditions. The limitation of this technique is that it can only provide the spectrum at this energy.

The predictions of reaction models giving cross sections relevant for the production of p-nuclei and of the nuclear properties required in such calculations can also be tested by using neutron and charged particle reactions on stable targets (Descouvemont and Rauscher 2006). Some of the accessible reactions are directly important in the nucleosynthetic processes while other experiments only fare as tests of the theoretical approaches (Kiss *et al.* 2008). The reactions can either be studied in online beam experiments detecting directly emitted γ-rays or particles. Another important type of experiment is that of activation (see, e.g., Gyürky *et al.* 2006, and references therein). A material sample is activated by neutron, proton, or α beams at the energy of interest and the long-term radiation is counted over an extended period of time. Alternatively, the amount of the nuclei produced by the activation can be measured by the very sensitive Accelerator Mass Spectrometry (AMS), which has become an important tool also for astrophysical measurements. It is especially well suited for treating neutron-induced reactions producing different isotopes of the same element.

The advent of radioactive ion beam (RIB) facilities in nuclear physics allows the study of the properties of and reactions with unstable nuclei (Käppeler *et al.* 1998, Thielemann *et al.* 2001b, Rauscher and Thielemann 2001). In addition to the few, already existing, smaller RIB facilities in Europe, Japan and the USA, one large-scale facility is under construction at GSI Darmstadt, Germany, and another large-scale facility has recently been funded in the USA, the Facility for Rare Isotope Beams (FRIB) at Michigan State University. Both the GSI FAIR (Facility for Antiproton and Ion Research) and FRIB will allow the production of highly unstable nuclides, both on the proton- and neutron-rich side of the chart of nuclides. For the first time, this will enable us to study nuclear properties of the p- and rp-processes directly, and also close to the r-process path. These investigations will largely improve the understanding of explosive nucleosynthetic processes.

Plasma physics experiments describe the properties of hot and thin plasmas. Utilizing data from laser-induced plasmas or nuclear testing allows drawing conclusions on the behavior of matter under conditions which are to a certain extent close to those found in stellar environments. Hydrodynamic simulations can also be validated against test cases drawn from experiments and terrestrial experience. However, large nuclear reaction networks involving highly unstable nuclei, extended stellar atmospheres with complicated mixing processes, or macroscopic amounts of matter at and beyond nuclear densities are only accessible by theoretical methods. The models have then to be tested against what we ultimately want to explain: astronomical data. The latter have almost exclusively been in the form of observations in the electromagnetic spectrum, starting from the ancient observations of visible light coming from the Sun and the stars, to modern satellite observatories also exploring other frequency ranges and studying emissions of compact objects, faint galaxies, accretion disks, quasars, and the echo of the Big Bang, the cosmic microwave background. It is amazing how much has already been learned about the structure and history of the whole Universe by just examining the faint light reaching the surface of our tiny planet. The upcoming new missions of ground- and space-based observatories guarantee an increasing inflow of data, securing the development of the related fields and ensuring that this research field stays exciting and is still able to provide new insights.

In addition to the observation in the electromagnetic spectrum, other means of obtaining astrophysically relevant information become increasingly important. Among those are measurements of cosmic rays, on the surface and in the atmosphere of Earth, as well as in low-earth orbits (Westphal *et al.* 1998, 2001). Such investigations provide insights regarding the particle flux in our solar system, originating from the Sun and from Galactic sources. In the future, we will be able to study another type of radiation in addition to the electromagnetic one: gravitational waves emitted, e.g., by neutron star mergers, black hole formation, and other interactions between highly massive or relativistic objects (Sathyaprakash and Schutz 2009). Even more important for studying nucleosynthesis and stellar evolution are the isotopic ratios found in certain meteoritic inclusions (Lugmair *et al.* 1983, Lewis *et al.* 1987, Hoppe and Zinner 2000). Using advanced chemical extraction methods, these data can be utilized to deduce the composition of the material in the early Solar System. Because of its growing importance for nucleosynthesis studies, the field is introduced in the separate Section 6.3.

## 6.2   Solar abundances

As might have become evident from the previous sections, the origin of the elements can only be understood in detail when also the physics of the nucleosynthesis sites is understood. Complete simulations of nucleosynthetic events are required which account for all data, not just those giving details of elemental or isotopic abundances. Neverthe-less, the determination of abundances remains central if one wants to study nucleo-synthesis.

By inspection of the absorption lines in stellar spectra it is possible to measure the contents of the stellar atmosphere. A theoretical model has to explain how many of those nuclides were inherited from the proto-cloud from which the star formed and what amount was produced in the central, nuclear burning regions of the star itself and brought up by convection.

Similar considerations apply to the observation of absorption and emission lines from other objects, such as supernova ejecta, planetary nebulae, and interstellar clouds. Some methods and results were already presented in Subsection 3.3.

If we want to understand the origin of elements on Earth, the abundances in the Sun have to be explained first because the planets and the Sun formed from the same interstellar cloud. Due to their low gravitation, the smaller planets subsequently lost those light elements which were not chemically bound in their crust whereas the Sun was able to retain more or less the original composition. In planets, physical and chemical fractionation processes then separated certain elements or isotopes and concentrated them in different regions, leading to the heterogeneous distribution found, e.g., in geological surveys. The solar composition is shown in Figure 13. As an example of how the solar composition affects nucleosynthesis studies is shown in Figure 12. In order to constrain the relative contributions of the s- and r-process, the solar abundances are used to represent the current composition of the local interstellar medium.

The accurate determination of solar abundances is, therefore, central to all investigations of nucleosynthesis. This is reflected in the recent commotion caused by a new study of solar abundances (Asplund et al. 2006), revising the previously widely used tables of Anders and Grevesse (1989). The new abundances are based on modern 3D model atmospheres (describing the region in the Sun where line absorption occurs). The content of elements beyond H and He was found to be lower by a factor of two compared to the previous study. This impacts all kinds of comparative nucleosynthesis studies but mainly those involving C, N, O which are the most abundant (apart from H,

He). The new abundances resolved previous problems regarding the consistency of solar abundances and those of the solar neighborhood. On the other hand, they challenge current models of the solar interior, especially regarding the comparison of predictions of the local sound speed (which is also dependent on the abundances of C, N, O, Ne) and helioseismological results.

## 6.3    Meteoritic inclusions

In addition to abundance determinations from stellar spectra, another way to obtain information about the composition of the early solar system as well as that of more distant environments has become increasingly important in the last years. Certain types of meteorites contain inclusions wherein the composition of the early pre-solar cloud is conserved. This enables one to also study some isotopic compositions which cannot be extracted from the solar spectrum. More surprisingly, some meteorites contain so-called **presolar grains** which are supposed to have been formed from material of other stars. The grains traveled through interstellar space with their inherent speed at formation and were incorporated into the protosolar cloud from which the Sun and its planets formed. Some of them survived the formation of the solar system and also the fall as a meteorite. This requires that their host material never experienced temperature above about 1000 K. Various types of meteorites, most prominently carbon-rich ones (carbonaceous chondrites) carry such nm to μm sized inclusions, which were incorporated into the presolar cloud and were since then shielded from any chemical or physical fractionation and mixing processes occurring during planet formation or in the Sun.

Literally being "stardust", presolar grains provide information not just on the isotopic composition of other stars which cannot be determined through their spectra. Additionally, they may show the composition of different layers of a star depending on their formation process. There are a number of excellent reviews on the topic (e.g. Nittler 2003, Clayton and Nittler 2004, Zinner *et al.* 2006, and the book by Lugaro 2005). Therefore, here we only summarize the most important aspects regarding types and origins of presolar grains and the methods to analyse them (see also Chapter xxx).

Presolar grains are foremost identified by their non-solar isotopic composition. They are subsequently classified by the mineral phase carrying the isotopes and by the isotopic ratios of certain elements (mainly C, N, O, Si, Al, Fe). Most abundant but the least understood are nanodiamonds. Best studied are the second most abundant SiC grains. Further phases include, in order of abundance, graphite, TiC, ZrC, MoC, RuC, FeC, Fe-Ni metal, $Si_3N_4$, corundum, spinel, hibonite, $TiO_2$.

The well-studied SiC grains can be subdivided into different classes according to the isotopic anomalies (relative to solar) they exhibit. The bulk of 90% is made of so-called **mainstream grains** which are thought to originate from AGB stars, showing almost pure s-process isotopic ratios. They are inferred to have formed in the winds of AGB stars or planetary nebulae. Therefore they are a snapshot of the surface composition of the star but AGB stars, contrary to other types of stars, have strong convection, carrying freshly synthesized nuclides from the burning zone deep inside the star to the surface (Section 4.5.1). Much of the recent progress on the nucleosynthetic details of the s-process in recent years is due to the analysis of presolar grains.

A small subclass of SiC grains, the so-called type X grains, contain a large $^{26}Mg/^{24}Mg$ ratio and large excess of $^{44}Ca$, pointing to a core-collapse supernova origin. The radioactive isotopes $^{26}Al$ and $^{44}Ti$ are concurrently produced only in such supernovae and decay to $^{26}Mg$ and $^{44}Ca$, respectively. The **SiC X grains** are thought to be formed in supernova ejecta. As these consist of a large fraction of the progenitor star, grains can also condense from material of inner layers or from a mixture of different layers of the star. There is some success in reproducing X grain compositions by mixing abundances predicted by current stellar models.

The origin of other types of grains is still debated and a unique identification with a site is not always possible. They could have been produced in supernovae, AGB stars, or novae.

The analysis of the content of a grain requires a combination of chemical and physical methods to separate the grain from the surrounding meteoritic material and to determine the isotopic abundances contained within. Luckily, mineral phases condensing in the vicinity of stars contain acid-resistant phases which can be separated from the meteorite by essentially dissolving everything else away. Until recently, this was the way to go but it has the disadvantage that other, less acid-resistant, presolar phases, e.g. silicates, may be lost in the process. Once the grain material has been isolated and concentrated, standard mass spectrometric methods (with AMS being the most sensitive) can be applied. A complete analysis of the presolar content, including more easily dissolvable materials, require new analytic methods, currently under development. Among those are Resonance Ionization Mass Spectrometry (RIMS), allowing the measurement of ppm-level trace elements in μm-sized grains with elimination of isobaric interferences, and even more promising the NanoSIMS, an ion microprobe with high sensitivity and high spatial resolution (Stadermann *et al.* 1999, Marhas *et al.* 2008). The NanoSIMS allows to study a grain in a slice of a meteorite and to obtain isotopic abundances with information on their location in the sample. This enables studies of layered grains in their meteoritic matrix and also grains made of easily dissolvable phases. This is superior to TEM (transmission electron microscope) analysis which requires ultra-thin samples cut with diamond knives and losing some depth information. Nevertheless, the TEM can sometimes be complementary to a NanoSIMS analysis.

Presolar grains open a new, promising window into the Universe by enabling us ``hands on'' analysis of non-solar, stellar matter. With improved preparation and analysis methods this line of research will remain important for many types of nucleosynthesis studies, even directly impacting the theory of stellar structure and evolution.

## 6.4 Galactical chemical evolution: putting it all together

Although the Sun is considered to exhibit a typical composition for the disk of the Galaxy, it has to be realized that its abundances are only a snapshot in time. As is evident from the discussion in the preceding sections, the Sun contains elements which have been produced in other stars or Galactic sites and the solar abundances (Figure 13) are only a snapshot of the composition of the interstellar medium. After the first stars in the galaxies lit up, numerous generations of stars have contributed to the elemental contents of the interstellar material from which new stars form. The current stars, on the other hand, are building up material which will be incorporated in future stellar generations. All this is illustrated by the fact that a gradient in metallicity is observed depending on the age of the star (see Subsection 3.3). Older stars contain less 'metals', i.e. elements other than H and He, because the interstellar medium from which they formed contained less. Therefore it becomes obvious that the interstellar medium in a galaxy becomes enriched with elements over time. The general picture is that of a cycle of matter within a galaxy as shown in Figure 14.

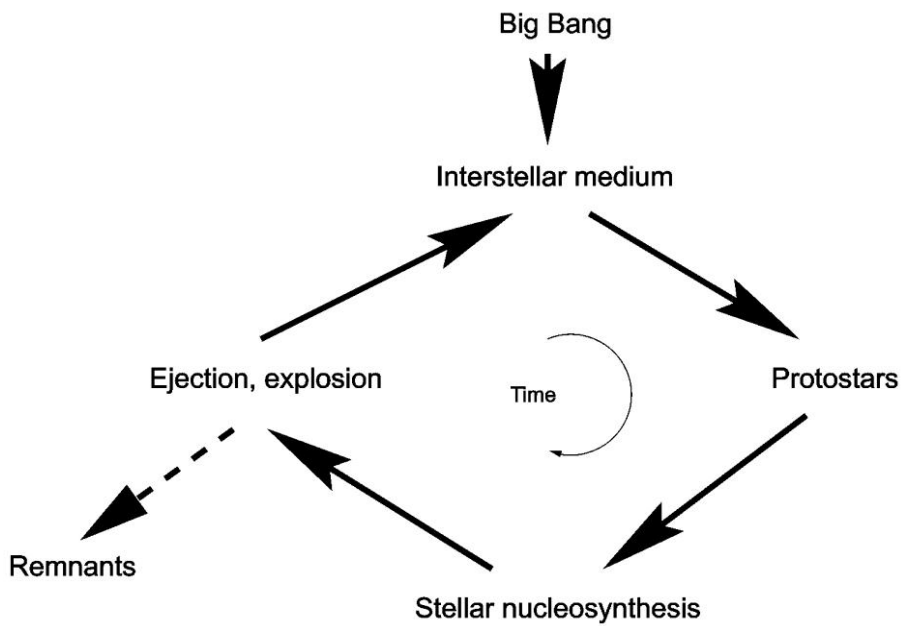

FIGURE 14. Schematic view of the cycle of matter in a galaxy.

The primordial galactic material is processed and reprocessed in star-forming regions many times. Indeed, a general enrichment can actually be found when comparing galaxies of different ages.

We do not know how many generations of stars have contributed to the solar abundances and how well the products were mixed into the proto-solar cloud. Therefore, a comparison with abundances in old stars allows to draw conclusions on the relevant processes. For example, recent observations in stars found in the halo of the Galaxy show that the relative r-process abundances are very similar to the ones in the Sun, although very much depleted (Sneden *et al.* 2000, Cayrel *et al.* 2001, Frebel *et al.* 2005). This indicates that the r-process seems to be robust, i.e. occurring under almost the same conditions and giving almost the same elemental yield in each event.

For a complete understanding of the chemical evolution of a galaxy it is necessary to integrate over the yields of all possible contributors. With the advent of advanced stellar models, galactic chemical evolution has become a field of its own, providing further constraints to the nucleosynthetic models. Many considerations enter, such as the amount and composition of ejecta per event, frequency of events, and mixing processes which distribute matter within a galaxy (Pagel 1997). Thus, all available knowledge is combined to reach an improved level of understanding. However, the young field of galactic chemical evolution still faces major difficulties due to the time-scales involved, the limitations in observations and models, and the impossibility of accurately dating stars and galaxies. One of the big questions is how material is mixed and transported. Nevertheless, recent promising trends in modeling galactic evolution might even provide constraints, e.g., on individual supernova models rather than only on global properties of SN II and SN Ia. The reason for this possibility is the fact that there is no instantaneous mixing of ejecta with the interstellar medium, and therefore early phases of galactic evolution can present a connection between low metallicity star observations and a single supernova event (Argast *et al.* 2000).

Summarizing, we want to emphasize again the tremendous achievements obtained over the last decades. Involving accurate and detailed studies it was made possible that, coming from more simple observations of the Sun and nearby stars, we are now in the position to study details of the origin of chemical elements and their isotopes on our planet as well as the evolution of their abundances in entire galaxies and in the early Universe. It is especially amazing to ponder this from the point of view that this knowledge was gathered without really or just barely leaving the surface of our planet.

Future efforts in nuclear physics and astronomy ensure that the stream of data will not be cut off and that we can improve our detailed knowledge not only of the origin of the elements but also of the position of our Galaxy, our planet, and ourselves within a vast, evolving Universe.